\documentclass[%
 reprint,
 amsmath,amssymb,
 aps,
superscriptaddress
]{revtex4-2}

\usepackage{float}
\usepackage{graphicx}
\usepackage{dcolumn}
\usepackage{bm}
\usepackage{quantikz}
\usepackage{braket}
\usepackage{booktabs}
\usepackage{listings}
\usepackage{xcolor}

\definecolor{codegreen}{rgb}{0,0.6,0}
\definecolor{codegray}{rgb}{0.5,0.5,0.5}
\definecolor{codepurple}{rgb}{0.58,0,0.82}
\definecolor{backcolour}{rgb}{0.95,0.95,0.92}

\lstdefinestyle{mystyle}{
    backgroundcolor=\color{backcolour},   
    commentstyle=\color{codegreen},
    keywordstyle=\color{magenta},
    numberstyle=\tiny\color{codegray},
    stringstyle=\color{codepurple},
    basicstyle=\ttfamily\footnotesize,
    breakatwhitespace=false,         
    breaklines=true,                 
    captionpos=b,                    
    keepspaces=true,                 
    numbers=left,                    
    numbersep=5pt,                  
    showspaces=false,                
    showstringspaces=false,
    showtabs=false,                  
    tabsize=4,
    language=Python
}
\newtheorem{Def}{Definition}

\newtheorem{The}{Theorem}

\definecolor{sergi}{rgb}{1,0,.5}

\begin{document}

\preprint{APS/123-QED}

\title{Experimental Realization of the Markov Chain\\Monte Carlo Algorithm on a Quantum Computer}

\author{Baptiste Claudon}\affiliation{Qubit Pharmaceuticals, Advanced Research Department, 75014 Paris, France}\affiliation{Sorbonne Universit\'e, LJLL, UMR 7198 CNRS, 75005 Paris, France}\affiliation{Sorbonne Universit\'e, LCT, UMR 7616 CNRS, 75005 Paris, France}\email{Contact authors: baptiste.claudon@qubit-pharmaceuticals.com,\\ jean-philip.piquemal@sorbonne-universite.fr}
\author{Sergi Ramos-Calderer}\affiliation{Centre for Quantum Technologies, National University of Singapore, Singapore}
\author{Jean-Philip Piquemal}\affiliation{Qubit Pharmaceuticals, Advanced Research Department, 75014 Paris, France}\affiliation{Sorbonne Universit\'e, LCT, UMR 7616 CNRS, 75005 Paris, France}

\date{\today}

\begin{abstract}
Quantum algorithms present a quadratically improved complexity over classical ones for certain sampling tasks. For instance, the Quantum Amplitude Estimation (QAE) algorithm promises to speed up the estimation of the mean of certain functions, given access to the quantum state corresponding to the probability distribution to be sampled from. Classically, samples are often obtained by simulating a Markov chain. In this work, we experimentally use encodings of Markov chains to prepare quantum states and run a quantum Markov Chain Monte Carlo (qMCMC) algorithm on Quantinuum System Model H2 and Quantinuum Helios quantum computers. We first demonstrate that it is possible to obtain accurate results by operating directly on the physical qubits. Crucially, to pave the way toward fault-tolerant execution, we also implement these algorithms using logical qubits protected by the Iceberg Quantum Error Detection (QED) code. By applying multiple rounds of mid-circuit QED during a random walk, we successfully catch and filter out hardware errors. These experiments demonstrate the feasibility of qMCMC on quantum hardware, bringing us closer to realizing quadratic quantum speedups with broad applications ranging from molecular modeling to machine learning and finance.
\end{abstract}

\maketitle

The pursuit of error-corrected protocols has seen steady improvement on quantum hardware. Contemporary quantum devices offer error rates at the point where complex subroutines, along with early active error mitigation techniques, are starting to be deployed with satisfactory results. While large-scale fault tolerance remains distant, this early stage offers the opportunity to gauge the development of this technology by pushing the limits of the available hardware. Still, this necessitates a twofold approach to be successful. Intimate knowledge of both the quantum device and middleware stack is essential to reach optimal performance. Moreover, breakthroughs at the algorithmic level reveal efficient protocols and decompositions that tighten the resources required from the machines in the first place. In this letter, we focus on this interplay. We perform a thorough test of the required subroutines for the quantum Markov Chain Monte Carlo (qMCMC) algorithm. Diverse encoding techniques are deployed and combined as per the recently proposed quantum encoding of the Metropolis-Hastings algorithm~\cite{claudon2025quantumcircuitsmetropolishastingsalgorithm} on ion trap quantum devices.

Building on Grover's work~\cite{grover}, Brassard introduced the Quantum Amplitude Estimation (QAE) algorithm~\cite{Brassard_2002}. Given access to quantum states encoding probability distributions, and a reversible algorithm to compute a function, the algorithm yields an estimate of its mean using quadratically fewer resources than the classical Monte Carlo method. Classically, samples from the distribution are often provided by implementing steps of a Markov chain until the state's distribution is close enough to its stationary measure. By analogy, it is possible to encode Markov chains into unitary operations. Such an encoding can be used to prepare the quantum state corresponding to the target stationary distribution, and therefore be used in Brassard's QAE method. 

In this work, we test several encodings of Markov chains on Quantinuum System Model H2 (H2-1 and H2-2) and Quantinuum Helios quantum computers~\cite{H2_Nature, H2_PRX, helios}. We first demonstrate a baseline full workflow on bare physical qubits by implementing a two-state Markov chain using the Linear Combination of Unitaries (LCU) technique. This encoding is used to sample the stationary measure and estimate an expectation value via the qMCMC algorithm. 

Second, to advance toward reliable execution, we implement the step operator of a four-state Simple Random Walk (SRW) using logical qubits protected by the Iceberg Quantum Error Detection (QED) code~\cite{Self_2024}. Applying successive rounds of mid-circuit QED on the Quantinuum Helios processor successfully suppresses unphysical hardware errors. This maps the trade-off between improving logical state accuracy and the cost of discarding flagged experimental runs.

Finally, we compare alternative encoding paradigms for the two-state system. We deploy Szegedy's quantization~\cite{szegedy} alongside a controlled-SWAP (cSWAP) encoding tailored for Metropolis-Hastings walks~\cite{hastings, claudon2025quantumcircuitsmetropolishastingsalgorithm}, highlighting how different algorithmic frameworks directly impact post-selection overhead. Building on this, we also scale our physical implementations to execute a lazy SRW on state spaces up to $\mathbb{Z}/64\mathbb{Z}$, demonstrating the hardware's ability to maintain high fidelity for much larger system sizes.

\section{Technical Background}

\subsection{Quantum Markov Chain \\ Monte Carlo}

Let $\mathbb S$ be a finite state space of size $n$. A Markov kernel is a matrix $P$ of positive numbers and size $n\times n$ such that $\sum_{y\in \mathbb S}P(x, y)=1$, for all $x\in \mathbb S$.
Throughout this work, we restrict to ergodic Markov chains, admitting a unique stationary distribution $\pi$ satisfying $\pi P=\pi$ and $\pi(x)>0$ for all $x\in \mathbb S$. We denote the coherent encoding of a probability distribution $\mu$ by $\ket\mu=\sum_{x\in \mathbb S}\sqrt{\mu(x)}\ket x$. In particular, $\ket\pi$ denotes the stationary state. Ref.~\cite{LevinPeresWilmer2006} provides a detailed exposition of the properties of such ergodic Markov kernels.

Quantum algorithms for Markov chains rely on embedding non-unitary stochastic operators into larger unitary operators. We here give a precise definition of such encoding~\cite{sünderhauf2023generalizedquantumsingularvalue}.

\begin{Def}
Let $U$ be a unitary acting on a Hilbert space $\mathcal H$, and let $\square:\mathbb C^n\to\mathcal H$ be a partial isometry. If $U$ is symmetric, then $(U, \square)$ is called a Symmetric Projected Unitary Encoding of an operator $A$ if $\square^\dag U\square=A$.
\end{Def}

These Symmetric Projected Unitary Encodings (SPUEs) naturally arise in quantum walk constructions. In practice, different implementations of $\square$ and $U$ correspond to different realizations of the same effective operator $A$. In this work we focus on experimentally comparing several such encodings.

Given a SPUE, one can construct a quantum walk operator whose spectrum is directly related to that of the encoded symmetric operator. This can be stated more precisely as follows.

\begin{Def}
Let $(U, \square)$ be a SPUE of an operator $A$. Its associated qubitized walk operator is 
\begin{equation}
\mathcal W=(2\square\square^\dag-1)U.
\end{equation}
\end{Def}

These SPUEs become the basis on which the quantum walk encoding is built. The following theorem underlies all subsequent algorithms~\cite{szegedy, sünderhauf2023generalizedquantumsingularvalue}.

\begin{The}
Let $(U, \square)$ be a SPUE of an operator $A$, with qubitized walk operator $\mathcal W$. Let $\lambda\in ]-1, 1[$ be an eigenvalue of $A$ with eigenvector $\ket v$. Define $\theta=\cos^{-1}(\lambda)$. Then $\mathcal W$ has eigenvalues $e^{\pm i\theta}$ with eigenvectors supported on the two-dimensional invariant subspace spanned by $\square\ket v$ and $U\square\ket v$. If $\lambda=\pm1$, then $\square\ket v$ is an eigenvector of $\mathcal W$ with the same eigenvalue. 
\end{The}

From here, the qubitized walk operator follows.
Given a Markov kernel $P$, define the quantum step operator
\begin{equation}
\square=\sum_{x\in \mathbb S}\ket x\ket{P(x, \cdot)}\bra x.
\end{equation}
For reversible chains with stationary distribution $\pi$, the discriminant matrix
\begin{equation}
\mathcal D(x, y)=\sqrt{\frac{\pi(x)}{\pi(y)}}P(x, y)
\end{equation}
is symmetric and shares the same spectrum as $P$. Using the SWAP operator $S$, the pair $(S, \square)$ is a SPUE of $\mathcal D$. This is referred to as Szegedy's quantization method~\cite{szegedy}. The associated qubitized walk operator therefore encodes the spectral properties of the original Markov chain. Importantly, the stationary state $\ket\pi$ satisfies $\mathcal D\ket\pi=\ket\pi$ and corresponds to the unique $1$ eigenstate of the walk operator in the range of $\square$. However, other partial isometries and unitary operators can also encode the discriminant matrix.

Let $\mathcal W$ denote the qubitized walk operator of a reversible ergodic chain. Since $\square\ket\pi$ is the unique eigenvector of $\mathcal W$ with eigenvalue $1$, it can be prepared using phase estimation~\cite{kitaev1995quantummeasurementsabelianstabilizer, nielsen}. Recall that given an eigenvector $\ket\psi$ of $\mathcal W$ with eigenvalue $e^{2\pi i\phi}$, $\phi\in \{k/2^t\}_{k=0}^{2^t-1}$, the phase estimation of $\mathcal W$ maps $\ket\psi\ket0$ to $\ket\psi\ket{2^t\phi}$. Together with post-selection, the algorithm can also serve to prepare eigenstates.

Ultimately, the target expectation value of the desired observable is extracted via amplitude estimation as follows.
Let $f:\mathbb S\to[0, 1]$ be an observable. The Monte Carlo algorithm aims at estimating 
\begin{equation}
\mathbb E_\pi(f)=\pi f=\sum_{x\in \mathbb S}\pi(x)f(x).   
\end{equation}
We assume access to $f$ through a unitary $O_f$ such that
\begin{equation}
O_f\ket{x, 0}=\sqrt{f(x)}\ket{x, 0}+\sqrt{1-f(x)}\ket{x, 1}.
\end{equation}
Applied to $\ket{\pi, 0}$, $O_f$ yields the state
\begin{equation}
\ket{\pi\cdot f}=\sum_{x\in \mathbb S}\sqrt{\pi(x)f(x)}\ket{x, 0}+\sqrt{\pi(x)(1-f(x))}\ket{x, 1}.
\end{equation}
In particular, $(2\ket{\pi\cdot f}\bra{\pi\cdot f}-1, \ket 0)$ is a SPUE of the symmetric operator
\begin{equation}
2\braket{0|\pi\cdot f}\braket{\pi\cdot f|0}-1,
\end{equation}
with eigenvalues $-1$ and $2\pi f-1$. Thus, applying the phase estimation algorithm to the qubitized walk operator of $(2\ket{\pi\cdot f}\bra{\pi\cdot f}-1, \ket 0)$ and initial state $\ket{\pi\cdot f}$ yields eigenvalue phases $\pm \cos^{-1}(2\pi f-1)$. Taking the cosine of either measured phase gives $2\pi f-1$.

\subsection{The Iceberg Quantum Error Detection Code}

The Iceberg code~\cite{Self_2024} is an $[[k+2, k, 2]]$ quantum error detection code that encodes $k$ logical qubits into $n=k+2$ physical qubits. We partition the physical qubits into a data register $D = \{1, \dots, k\}$ and a parity register $P = \{t, b\}$, corresponding to a ``top'' and ``bottom'' parity qubit. 

The logical subspace is defined by two global parity checks.

\begin{Def}
Let $S_X = \bigotimes_{i \in D \cup P} X_i$ and $S_Z = \bigotimes_{i \in D \cup P} Z_i$ be global physical Pauli operators. The Iceberg code space $\mathcal{C}$ is the $2^k$-dimensional simultaneous $+1$ eigenspace of $S_X$ and $S_Z$. A state $\ket{\overline{\psi}}$ is a valid logical state if and only if $S_X\ket{\overline{\psi}} = \ket{\overline{\psi}}$ and $S_Z\ket{\overline{\psi}} = \ket{\overline{\psi}}$.
\end{Def}

Logical Pauli operators in this code can be taken to be weight-2 Pauli operators acting on a data and a parity qubit.

\begin{Def}
For any data qubit index $j \in D$, the single-qubit logical Pauli operators are
\begin{equation}
\overline{X}_j = X_t X_j \quad \text{and} \quad \overline{Z}_j = Z_b Z_j.
\end{equation}
\end{Def}

By construction, these weight-2 operators naturally commute with both global parity checks. This mathematical property guarantees that applying any sequence of valid logical gates will strictly preserve the system's presence within the error-free code space.

Crucially, for distinct logical qubits $i, j \in D$, the product of their logical Paulis reduces to a physical weight-2 operator acting solely on the data register as
\begin{equation}
\overline{X}_i \overline{X}_j = (X_t X_i)(X_t X_j) = X_i X_j.
\end{equation}
Therefore, continuous logical two-qubit rotations, such as $\exp(-i\theta \overline{X}_i \overline{X}_j)$, are mathematically identical to their physical counterparts $\exp(-i\theta X_i X_j)$ and can be implemented directly via standard two-qubit entangling gates. We execute these using partially Fault-Tolerant (pFT) gadgets; by minimizing multi-qubit overhead and circuit depth, pFT implementations yield superior performance on current noisy hardware.

To identify hardware-induced errors during execution, we exploit that any single physical Pauli error anti-commutes with $S_X$ or $S_Z$ (since $d=2$). We enforce evolution within $\mathcal{C}$ by periodically measuring these stabilizers mid-circuit using a pFT approach. Specifically, each syndrome extraction applies a basis change to the physical qubits, sequentially applies CNOT gates targeting a single auxiliary qubit, measures the ancilla to extract the parity, and reverts the basis. If a $-1$ eigenvalue is recorded, an error is deterministically flagged. By discarding these corrupted runs, we project the remaining statistical ensemble back into the valid logical subspace before the next algorithmic step.

\section{Results}

\subsection{Full-Workflow Demonstration: State Preparation and Amplitude Estimation}

As a baseline demonstration of the end-to-end quantum algorithm workflow, we consider a simple random walk on a two-state space $\mathbb{S}=\{0, 1\}$. The walk is governed by a Markov kernel $P$ of the form:
\begin{equation}
P=\begin{pmatrix}
1-\delta& \delta\\
\delta& 1-\delta
\end{pmatrix}, \quad \delta>0,
\end{equation}
which admits a uniform stationary measure $\pi = \begin{pmatrix} 0.5 & 0.5 \end{pmatrix}$. For this initial set of experiments, we set the transition parameter to $\delta=1/4$. We evaluate the performance of state preparation and amplitude estimation for this system on both the H2-1 and Helios quantum processors.

\paragraph*{Linear Combination of Unitaries.}
To encode the Markov kernel, we use the LCU technique to express $P$ as $P=(1-\delta)I+\delta X$, where $I$ is the identity matrix and $X=\ket{0}\bra{1}+\ket{1}\bra{0}$. We implement the corresponding block encoding via a two-qubit unitary $U$ defined as:
\begin{equation}
\begin{quantikz}
\lstick{$a$:}   &\gate[2]{U}    &\\
\lstick{$x$:}   &               &
\end{quantikz}
=
\begin{quantikz}
\lstick{$a$:}   &\gate{R_Y(\theta)} &\ctrl{1}   &\gate{R_Y(-\theta)}    &\\
\lstick{$x$:}   &                   &\targ{}    &                       &
\end{quantikz},
\end{equation}
where $\theta=\cos^{-1}\left(\sqrt{1-\delta}\right)=\pi/6$. Here, $a$ designates the ancilla qubit and $x$ the state space qubit. The pair $(U, \ket{0})$ forms a SPUE of $P$. The corresponding qubitized walk operator is $\mathcal{W}=ZU$ (where $Z$ acts on $a$), which can be straightforwardly controlled by an auxiliary qubit $c$, as depicted in Figure~\ref{fig:controlled_qubitized_lcu}.

\begin{figure}[H]
\centering
\begin{quantikz}
\lstick{$c$:}	&				&\ctrl{1}	&				&\ctrl{1}	&\\
\lstick{$a$:}   &\gate{R_Y(\theta)}	&\ctrl{1}	&\gate{R_Y(-\theta)}	&\gate{Z}	&\\
\lstick{$x$:}   &				&\targ{}	&				&		&
\end{quantikz}
\caption{Controlled-$\mathcal{W}$ circuit. $c$ labels the control qubit, $a$ the ancilla qubit, and $x$ the state space qubit.}
\label{fig:controlled_qubitized_lcu}
\end{figure}

Because our specific choice of transition parameter $\delta=1/4$ yields a rotation angle of $\theta=\pi/6$, applying $\mathcal{W}^3$ exactly implements the reflection $2\ket{\pi}\bra{\pi}-1$. We utilize this block encoding to prepare the stationary state $\ket{\pi}$ according to the circuit in Figure~\ref{fig:state_preparation_circuit}.

\begin{figure}[H]
\centering
\begin{quantikz}
\lstick{c:$\ket{0}$}  &\gate{H}   &\ctrl{1}                   &\gate{H}   &\meter{}\\
\lstick{$a:\ket{0}$}  &           &\gate[2]{\mathcal{W}^3}    &\rstick{$\ket{0}$}\\
\lstick{$x:\ket{0}$}  &           &                           &
\end{quantikz}
\caption{State preparation circuit. If the control qubit $c$ is measured in state $\ket{0}$, the ancilla $a$ is successfully projected to $\ket{0}$ and the target $x$ is prepared in the stationary state $\ket{\pi}$.}
\label{fig:state_preparation_circuit}
\end{figure}

Table~\ref{tab:lcu_state_preparation} summarizes the state preparation sampling statistics. Out of $10^4$ initial shots, both hardware platforms successfully conditioned the state and effectively reproduced the expected uniform distribution over the valid samples. 

\begin{table}[H]
\centering
\begin{tabular}{lccc}
\toprule
& \textbf{$\ket{0}$} & \textbf{$\ket{1}$} & \textbf{Successes} \\
\midrule
\textbf{Expected} & 2500 & 2500 & 5000 \\
\textbf{H2-1}     & 2586 & 2447 & 5033 \\
\textbf{Helios}   & 2321 & 2459 & 4780 \\
\bottomrule
\end{tabular}
\caption{Sampled outputs for the stationary state preparation initialized at $\ket{0}$ ($10^4$ total shots) using the LCU method. The number of successes corresponds to the instances where the control qubit was measured in the $\ket{0}$ state.}
\label{tab:lcu_state_preparation}
\end{table}

Following successful state preparation, we applied quantum amplitude estimation to compute the expected value of the observable function $f(x)=\delta_1(x)$, which theoretically yields $\mathbb{E}_\pi(f)=1/2$. The measurement distributions, summarized in Table~\ref{tab:lcu_monte_carlo}, demonstrate strong agreement with the target expectation value on both devices.

\begin{table}[H]
\centering
\begin{tabular}{lccc}
\toprule
\textbf{Measured $\mathbb{E}_\pi(f)$:}   &  \textbf{0}   & \textbf{0.5} & \textbf{1} \\
\midrule
\textbf{Expected}               & 0             & 500 & 0 \\
\textbf{H2-1}                   & 12            & 444 & 29 \\
\textbf{Helios}                 & 8             & 464 & 16 \\
\bottomrule
\end{tabular}
\caption{Measured expectation values for $\mathbb{E}_\pi(f)$ out of $10^3$ shots via the LCU method. The state preparation step succeeded $485$ times on H2-1 and $488$ times on Helios.}
\label{tab:lcu_monte_carlo}
\end{table}

\subsection{Error-Corrected Step Operator for a Four-State Random Walk}

In this section, we present an error-corrected implementation of the quantum step operator $O$ for an SRW on a four-state space ($\mathbb{S} = \mathbb{Z}/4\mathbb{Z}$). The step operator is defined such that its action on the walker yields $O\ket{x, 0} = \frac{1}{\sqrt{2}}\ket{x, x} + \frac{1}{2}\ket{x, x+1} + \frac{1}{2}\ket{x, x-1}$. When the initial register is set to $x=0$, this ideal evolution leaves the target register $y$ at state $0$ with a probability of $1/2$, and transitions to states $1$ and $3$ with a probability of $1/4$ each. The initial register $x$ should remain unchanged. In Appendix~\ref{app:lsrw_circuits}, we describe how to implement the quantum step operator for a general SRW on $\mathbb{Z}/k\mathbb{Z}$ for arbitrary state space sizes $k$.

We executed the corresponding circuits on the Helios quantum processor to compare two primary encoding strategies: a direct implementation on bare physical qubits, and a fault-tolerant implementation utilizing 4 logical qubits encoded into 6 physical qubits via the Iceberg QED code. To evaluate the impact of active error detection, the logically encoded circuits were tested with $0$, $1$, $3$, and $5$ rounds of mid-circuit QED. With each round requiring two additional auxiliary qubits for syndrome extraction, these configurations utilize a total of $6$, $8$, $12$, and $16$ physical qubits, respectively. In all encoded experiments, regardless of the number of mid-circuit rounds, we applied a final layer of error mitigation by extracting the expected parity checks from $Z$-basis measurements at the end of the circuit and post-selecting the results. All the circuits executed in this subsection are given in the Appendix~\ref{app:compiled_circuits}.

\begin{figure*}[t]
\centering
\includegraphics[width=\textwidth]{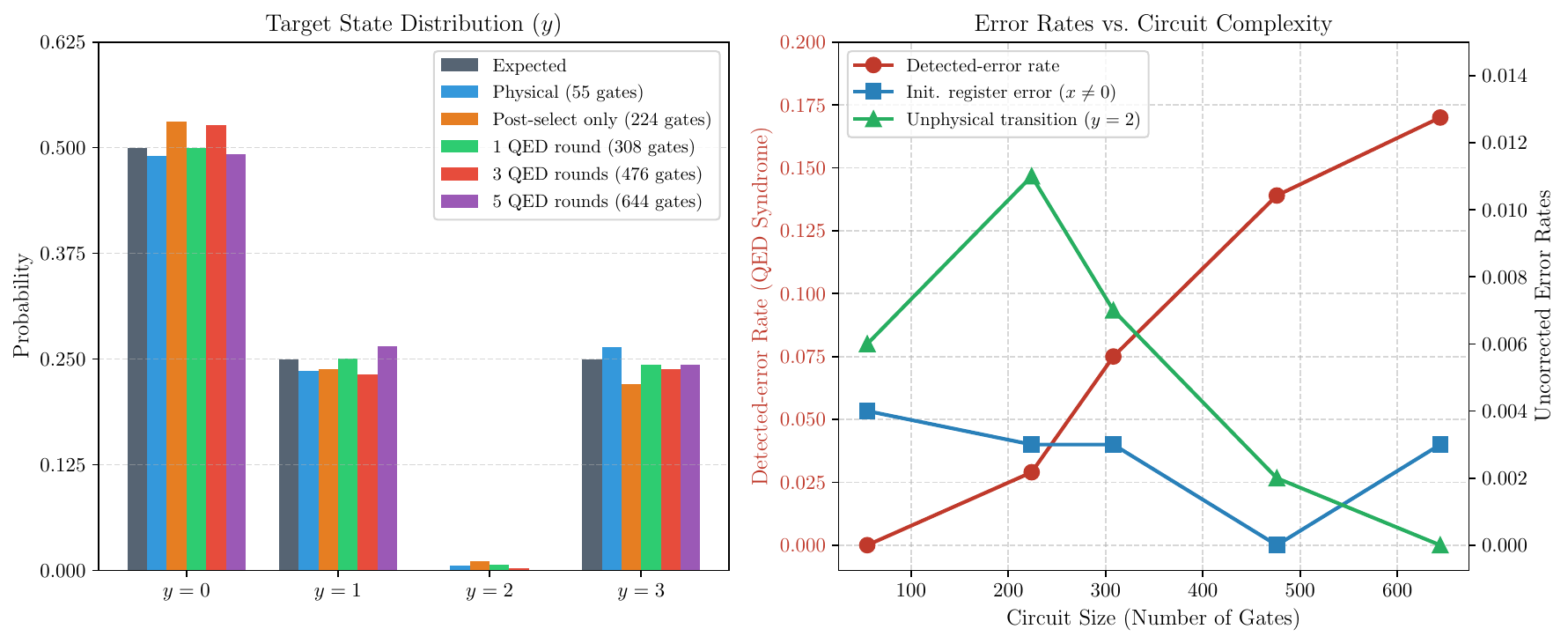}
\caption{Experimental results for the SRW step operator on Helios, based on $10^3$ shots per configuration. \textbf{Left:} Sampled state distributions ($y=0,1,2,3$) shown as normalized probabilities after post-selection. The baseline physical qubit implementation is compared against Iceberg QED logical encodings with varying rounds of mid-circuit error detection. \textbf{Right:} The trade-off between the detected-error rate (fraction of runs discarded due to QED syndrome measurements) and the uncorrected error rates (both the initial register error $x \neq 0$ and the unphysical transition to $y=2$) as a function of the total circuit size. Circuit sizes scale from 55 gates for the bare physical implementation up to 644 gates for the 5-round QED implementation.}
\label{fig:qec_srw}
\end{figure*}

Consequently, experimental runs can be discarded due to two distinct types of detected errors. First, in the logically encoded circuits, runs are discarded if an error is flagged by the QED syndrome measurements. Second, if an outcome shows that the initial state register $x$ is no longer $0$, we know an uncorrected error has occurred. The sampled target state ($y$) distributions for valid runs, extracted from a total of $10^3$ shots per configuration, alongside these two error metrics and their relationship to the total circuit size (which scales from 55 up to 644 operations), are summarized in Figure~\ref{fig:qec_srw}.

\subsection{Scaling to Larger State Spaces and Alternative Encodings}

Building on the baseline two-state demonstrations, we scaled the physical implementations to larger state spaces. Specifically, we implemented the step operator for a lazy simple random walk on $\mathbb{Z}/n\mathbb{Z}$ for dimensions $n=4, 8, 16, 32,$ and $64$. The compilation of these larger random walks, described in Appendix~\ref{app:lsrw_circuits}, uses the Draper Adder~\cite{draper2000additionquantumcomputer} method with quantum Fourier transforms, with circuit sizes scaling from 83 gates for $n=4$ up to 399 gates for the $n=64$ system. Despite these increasing circuit depths, the hardware retained high fidelity up to 64 states. Post-selected on the second register being $y=0$, the measured fidelities of $O^\dag SO\ket{0, 0}$ relative to the ideal target distribution were $0.989, 0.984, 0.968, 0.979,$ and $0.984$ for the respective system sizes. Furthermore, the experimental post-selection rates remained stable across all dimensions, consistently staying between $34.8\%$ and $37.1\%$, closely matching the theoretical expectation of $37.5\%$.

While the LCU method provides a direct approach to encoding the Markov kernel, quantum algorithms offer alternative paradigms for constructing the required SPUE. We evaluated these alternatives by summarizing the hardware performance of three distinct encoding strategies for the two-state system: Szegedy's quantization, a controlled-SWAP (Metropolis-Hastings) formulation, and a dual space encoding method. The Szegedy approach successfully prepared the uniform stationary state, recovering the expected distribution with an experimental post-selection success rate of $49.3\%$, closely matching the theoretical $50\%$ probability. In contrast, the controlled-SWAP method circumvented this sampling overhead entirely; it achieved a theoretical $100\%$ acceptance rate, reflected experimentally by $96\%$ of the state preparation shots successfully measuring in the correct algorithmic phase. Finally, we implemented the dual space method, yielding state preparation fidelities of $68.8\%$, $65.0\%$, and $70.7\%$ on the bare physical qubits of the Quantinuum System Model H2 (H2-1 and H2-2), and Quantinuum Helios processors, respectively. Such fidelities are in agreement with the announced $99.9\%$ and $99.921\%$ two-qubit gate fidelities of the Quantinuum System Model H2 (H2-1 and H2-2), and Quantinuum Helios (respectively). Comprehensive details, including quantum circuits, complete sampling statistics for the Szegedy and controlled-SWAP methods, and an extended analysis of the dual space encoding, are provided in Appendix~\ref{app:extended_results} and Appendix~\ref{app:data}.

\section{Discussion}

In this work, we have extensively explored the relationship between quantum hardware capabilities and algorithmic compilation for the quantum Markov Chain Monte Carlo algorithm. By demonstrating the full workflow of state preparation and amplitude estimation directly on bare physical qubits, we established a strong baseline. The Quantinuum System Model H2 and Quantinuum Helios processors exhibited sufficient native fidelity to successfully execute the required block encodings and phase estimation routines without active error correction. 

To push beyond the limitations of physical qubits and pave the way toward fault tolerance, we integrated the Iceberg Quantum Error Detection (QED) code into a four-state random walk. The results from the Helios processor clearly illustrate the power of error detection. Crucially, we observe that applying more than three rounds of mid-circuit error detection enables the logically encoded circuits to outright outperform the bare physical baseline in terms of uncorrected errors. Remarkably, by scaling the logical circuits up to five rounds of syndrome measurements, we completely eliminated the unphysical transitions to the $y=2$ state. This achievement highlights a fundamental dynamic in early-stage quantum error correction. We can effectively suppress hardware-induced leakage and boost algorithmic fidelity to near-ideal theoretical levels, at the direct cost of discarding a larger fraction of experimental runs due to flagged syndromes.

Furthermore, the hardware's versatility is not limited to a single algorithmic approach. As extensively detailed in the Appendix, our exploration of alternative encoding paradigms (including Szegedy's quantization, controlled-SWAP, and dual space methods) confirms that algorithmic choice heavily dictates post-selection overhead and native circuit depth. Simultaneously, our successful scaling of the lazy simple random walk up to 64 states demonstrates that trapped ions can reliably maintain high fidelity through complex operations such as the Draper Adder and quantum Fourier transforms. 

It is crucial to contextualize these hardware benchmarks within the broader landscape of quantum MCMC algorithms. In practice, classical MCMC methods are rarely used in their vanilla form; instead, they are often heavily improved through empirical modifications and heuristics to accelerate convergence. To complement these techniques in the near term, heuristic approaches such as quantum-enhanced MCMC have been successfully deployed on superconducting processors to serve as quantum proposal generators for classical MCMC loops~\cite{Layden_2023}. While such hybrid techniques bypass deep coherent circuit requirements and demonstrate empirical improvements in convergence, they lack the provable quadratic speedups of phase-estimation-based methods. Conversely, realizing these rigorous quadratic speedups presents a formidable challenge due to the massive hardware overheads associated with full fault tolerance. As recent analyses have emphasized, the constant-factor slowdowns introduced by quantum error correction mean that QAE-based algorithms will require dramatic improvements in code efficiency and logical clock speeds before they can achieve practical time-to-solution advantages over classical counterparts~\cite{Babbush_2021}. By empirically mapping the trade-offs between active error detection overheads and algorithmic fidelity on state-of-the-art trapped-ion hardware, our experiments directly confront the strict requirements necessary to eventually bridge this gap.

Looking forward, the natural next step is to merge these complementary approaches. The successful deployment of mid-circuit QED proves that logical encoding is viable on current devices, while our 64-state implementations prove that complex algorithmic scaling is achievable. As device coherence and qubit counts continue to scale, combining robust logical encodings with scalable operations will enable much larger error-corrected experiments. Ultimately, this will allow for the execution of deep, fully fault-tolerant qMCMC algorithms on complex, high-dimensional target distributions, moving us closer to realizing the quadratic speedups promised by quantum amplitude estimation.

\section{Acknowledgments}
This work has received funding from the European Research Council (ERC) under the European Union's Horizon 2020 research and innovation program (grant agreement No 810367), project EMC2 (JPP); and from the HQI program (JPP). Computations on the Quantinuum System Model H2 machines have been made possible through the Quantinuum Startup Partner Program. This project is supported by the National Research Foundation, Singapore through the National Quantum Office, hosted in A*STAR, under its Advanced Quantum Computing Infrastructure Funding Initiative (S25Q8D9001)

\section{References}
\bibliography{apssamp}

\onecolumngrid
\appendix

\section{Extended Experimental Results and Alternative Encodings}
\label{app:extended_results}

All hardware experiments presented in this work were executed through Quantinuum's \emph{nexus} platform~\cite{quantinuum_nexus}. Circuit compilation and submission were handled using \emph{pytket}~\cite{pytket} for the Quantinuum System Model H2 deployments and \emph{guppylang}~\cite{koch2025guppy} for the Quantinuum Helios processor. In this appendix, we provide the full theoretical constructions and compiled quantum circuits for the alternative Markov chain encodings summarized in the main text.

\subsection{Szegedy's Quantization}
\label{app:szegedy}

As an alternative to the Linear Combination of Unitaries (LCU) approach, the transition kernel $P$ can be encoded following Szegedy's quantization procedure~\cite{szegedy}. For a two-state system with transition probability $\delta=1/4$, this method requires a unitary operator $O$ satisfying:
\begin{equation}
O\ket{0, 0}=\sqrt{1-\delta}\ket{0, 0}+\sqrt{\delta}\ket{0, 1},
\end{equation}
and 
\begin{equation}
O\ket{1, 0}=\sqrt{\delta}\ket{1, 0}+\sqrt{1-\delta}\ket{1, 1}.
\end{equation}
A specific gate-level implementation of $O$ is provided in Figure~\ref{fig:def_qsteop}. Using the SWAP operator $S$, the tuple $(S, O\ket{0}, O\ket{0})$ forms a Symmetric Projected Unitary Encoding (SPUE) of $P$. The corresponding qubitized walk operator is $\mathcal W=OZO^\dag S$, which can be used as in Figure~\ref{fig:state_preparation_circuit} to prepare the stationary state $\ket{\pi}$. The full measurement distributions from this state preparation across the H2-1 and Helios devices are reported in Table~\ref{tab:szegedy_state_preparation}.

\begin{figure*}[h]
\begin{equation}
\begin{quantikz}
\lstick{$x$:}   &\gate[2]{O}    &\\
\lstick{$y$:}   &               &
\end{quantikz}
=
\begin{quantikz}
\lstick{$x$:}   &\targ{}     &           &\ctrl{1}           &\targ{}    &\ctrl{1}           &           &              &\\
\lstick{$y$:}   &\gate{S}    &\gate{H}   &\gate{R_Z(-\pi/3)} &           &\gate{R_Z(-2\pi/3)} &\gate{H}  &\gate{S^\dag} &
\end{quantikz}
\end{equation}
\caption{Quantum circuit implementing the unitary $O$ in Szegedy's quantization procedure.}
\label{fig:def_qsteop}
\end{figure*}

\begin{table}[h]
\centering
\begin{tabular}{lcccc}
\toprule
\textbf{Target State $(x,y)$:} & \textbf{$(0, 0)$} & \textbf{$(0, 1)$} & \textbf{$(1, 1)$} & \textbf{$(1, 0)$} \\
\midrule
\textbf{Expected} & 1875 & 625 & 1875 & 625 \\
\textbf{H2-1}   & 1711 & 694 & 1779 & 625 \\
\textbf{Helios} & 2054 & 543 & 1743 & 662 \\
\midrule
\textbf{Marginal State $x$:} & \multicolumn{2}{c}{\textbf{$0$}} & \multicolumn{2}{c}{\textbf{$1$}} \\
\midrule
\textbf{Expected} & \multicolumn{2}{c}{2500} & \multicolumn{2}{c}{2500} \\
\textbf{H2-1}   & \multicolumn{2}{c}{2405} & \multicolumn{2}{c}{2404} \\
\textbf{Helios} & \multicolumn{2}{c}{2597} & \multicolumn{2}{c}{2405} \\
\bottomrule
\end{tabular}
\caption{State preparation using Szegedy's method ($10^4$ total shots). The state preparation succeeded $4809$ times on H2-1 and $5002$ times on Helios, closely aligning with the theoretical success rate of $0.5$.}
\label{tab:szegedy_state_preparation}
\end{table}

\subsection{Controlled-SWAP Encoding}
\label{app:cswap}

Another paradigm for encoding $P$ is to frame it as a Metropolis-Hastings kernel. In this framework, the proposal kernel is $T=X$ and each proposed move is accepted with probability $\delta$. For our two-state implementation, we define the acceptance operator $O_A=\exp\left(i\pi Y/6\right)$ and the proposal operator $O_T$ as:
\begin{equation}
\begin{quantikz}
\lstick{$x$:}   &\gate[2]{O_T}  &\\
\lstick{$y$:}   &               &
\end{quantikz}
=
\begin{quantikz}
\lstick{$x$:}   &\ctrl{1}   &           &\\
\lstick{$y$:}   &\targ{}    &\targ{}    &
\end{quantikz}.
\end{equation}
Here, $(S^c, O_A O_T\ket{0, 0})$ forms the SPUE of $P$, where $S^c$ denotes the controlled-SWAP (cSWAP) operation. We prepare the state $O_A O_T\ket{\pi, 0, 0}$ and measure both the $x$ register and the phase auxiliary qubit. Because the cSWAP method theoretically guarantees an acceptance rate of $1$, it completely circumvents the sampling overhead present in Szegedy's approach. Table~\ref{tab:cswap_measurements} confirms this high empirical acceptance rate across hardware platforms.

\begin{table}[h]
\centering
\begin{tabular}{lccc}
\toprule
\textbf{Target State $x$:} & \textbf{$0$} & \textbf{$1$} & \textbf{Phase $0$ measurements} \\
\midrule
\textbf{Expected} & 5000 & 5000 & 10000 \\
\textbf{H2-1}   & 4948 & 4732 & 9680 \\
\textbf{Helios} & 5013 & 4738 & 9751 \\
\bottomrule
\end{tabular}
\caption{State preparation using the cSWAP encoding over $10^4$ shots. The number of phase $0$ measurements denotes the successful runs of the phase estimation step.}
\label{tab:cswap_measurements}
\end{table}

\subsection{Dual Space Quantum Markov Chain Monte Carlo}
\label{app:dual_space}

For any Metropolis-Hastings kernel, utilizing only $O_T$ and $O_A$, it is possible to encode the quantum step operator of a Markov process on the set of possible state pairs. The first marginal of this encoded process directly corresponds to the Metropolis-Hastings Markov chain. Operating in this dual space avoids the costly overhead required to track acceptance and rejection probabilities via auxiliary registers. 

The general construction of the corresponding qubitized walk operator $\mathcal W$ is detailed in Ref.~\cite{claudon2025quantumcircuitsmetropolishastingsalgorithm}. To validate the hardware execution of this dual space formulation using $T=X$ and $O_A=e^{i\pi Y/4}$, we designed a verification experiment based on the stationary state. We define a unitary $V$ such that the prepared state $V\ket{0}$ is an exact $1$-eigenstate of the walk operator $\mathcal W$. Consequently, the state should remain entirely invariant under the application of $\mathcal W$. 

Written in the computational basis (base 10 representation), this target eigenstate is:
\begin{equation}
V\ket{0} = \frac{1}{\sqrt{8}}\left(\ket{10}+\ket{12}+\ket{18}+\ket{20}+\ket{42}+\ket{44}+\ket{50}+\ket{52}\right).
\end{equation}

\begin{figure*}[h]
    \centering
    \includegraphics[width=\linewidth]{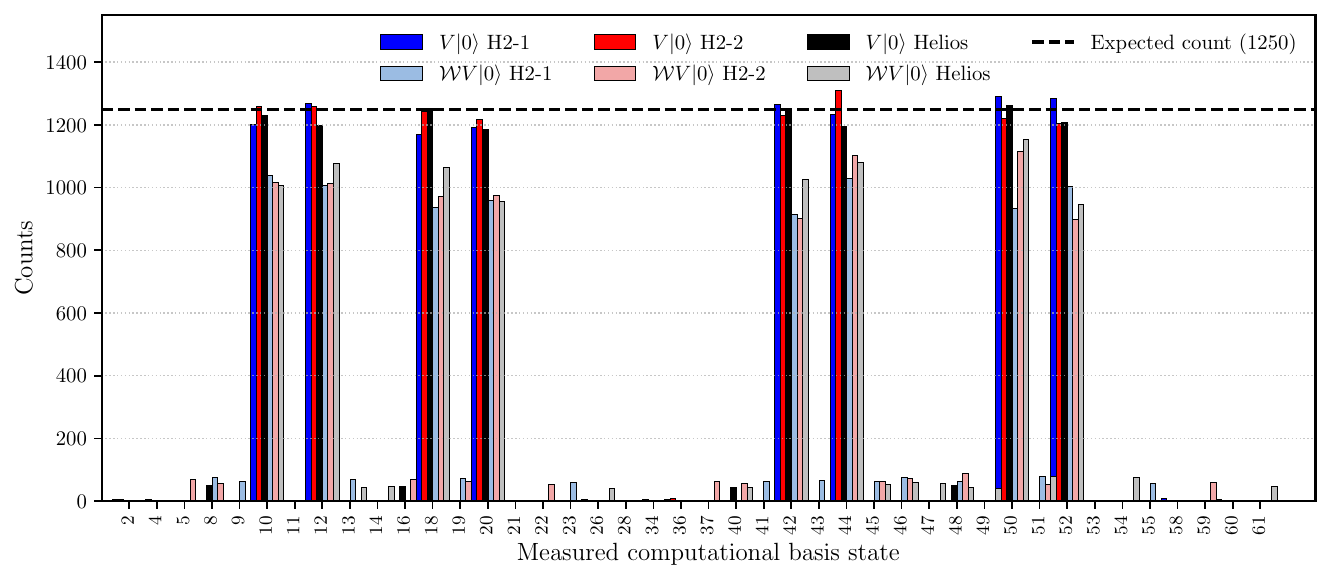}
    \caption{Measurement statistics from $10^4$ shots of $V\ket{0}$ and $\mathcal W V\ket{0}$ on Quantinuum's H2-1, H2-2, and Helios devices. Because $V\ket{0}$ is a $1$-eigenvector of the walk operator $\mathcal W$, the output probability distribution remains strictly invariant after the application of $\mathcal W$.}
    \label{fig:histogram}
\end{figure*}

We measured the computational basis statistics of both the initially prepared eigenstate $V\ket{0}$ and the evolved state $\mathcal W V\ket{0}$. The sampled distributions, presented in Figure~\ref{fig:histogram}, demonstrate strong agreement before and after the application of the walk operator, visually confirming the eigenstate property. 

To rigorously quantify this invariance, we directly measured the state overlap $|\braket{0|V^\dagger\mathcal WV|0}|^2$ by applying the reverse preparation circuit $V^\dag$ after the walk step and tallying the number of observed $0$ outcomes across the entire register. The raw hardware counts for this overlap fidelity measurement are detailed in Table~\ref{tab:overlaps}.

\begin{table}[h]
\centering
\begin{tabular}{lcccc}
\toprule
\textbf{Device} & \textbf{H2-1} & \textbf{H2-2} & \textbf{Helios} & \textbf{Expected} \\
\midrule
\textbf{Target state $0$ outcomes} & 688 & 650 & 707 & 1000 \\ 
\bottomrule
\end{tabular}
\caption{Overlap measurement evaluating the invariance of the dual space stationary state. The values represent the number of all-zero outcomes observed when measuring the total register for $V^\dag \mathcal WV\ket{0}$, based on $1000$ total shots per device.}
\label{tab:overlaps}
\end{table}

\section{Quantum Step Operator of the lazy Simple Random Walk}\label{app:lsrw_circuits}

To execute the quantum step operator $O$ for a lazy simple random walk on $n=2^N$ states, we must map the initial state $\ket{x}\ket{0}$ to $\frac{1}{\sqrt{2}}\ket{x}\ket{x} + \frac{1}{2}\ket{x}\ket{x+1} + \frac{1}{2}\ket{x}\ket{x-1} \pmod{2^N}$. This is achieved in two stages.

\paragraph{Coin State Preparation.} 
We first prepare the allowed step transitions $\Delta \in \{0, +1, -1\}$ inside the target register $y$ using a two's complement binary representation. Indexing from the least significant bit ($y_0$) to the most significant bit ($y_{N-1}$), the sequence requires:
\begin{enumerate}
    \item A Hadamard gate on $y_0$.
    \item A controlled-Hadamard (CH) gate from $y_0$ to $y_1$.
    \item A cascade of CNOT gates where each $y_i$ acts as the control for $y_{i+1}$, for $i$ from $1$ up to $N-2$.
\end{enumerate}
This sequence natively constructs the state $\frac{1}{\sqrt{2}}\ket{0\dots00} + \frac{1}{2}\ket{0\dots01} + \frac{1}{2}\ket{1\dots11}$, encoding the integer offsets $0$, $+1$, and $-1 \pmod{2^N}$ with their exact corresponding random walk amplitudes.

\paragraph{Draper Addition.}
To finalize the step operator, we coherently add the starting position $x$ to the transition offset $\Delta$ using the Draper adder~\cite{draper2000additionquantumcomputer}. By applying a Quantum Fourier Transform (QFT) to the $y$ register, followed by a sequence of controlled-phase gates from $x$ to $y$, and concluding with an inverse QFT, the adder evaluates the in-place modular arithmetic $\ket{x, y} \mapsto \ket{x, x+y \pmod{2^N}}$. Because the phase rotations naturally wrap around $2\pi$, this implements the periodic boundary conditions of the random walk without requiring any auxiliary qubits.

\section{Error-Corrected Step Operator: Compiled Circuits}\label{app:compiled_circuits}

In this appendix, we present the explicitly compiled quantum circuits used to execute the step operator $O$ for the four-state random walk. The circuits are expressed in Guppy, a Python-based quantum programming language, and are compiled directly into the native gate set of the Quantinuum Helios hardware. 

The Quantinuum architecture supports the following native operations:
\begin{itemize}
    \item \textbf{Z-Rotation:} \texttt{rz(q, $\theta$)} applies a local phase rotation $R_z(\theta) = \exp(-i \frac{\theta}{2} Z)$. These are implemented purely in software via frame-tracking and thus have zero physical execution time and zero error.
    \item \textbf{Phased-X:} \texttt{phased\_x(q, $\theta$, $\phi$)} applies a rotation by angle $\theta$ around an axis in the $XY$-plane defined by the azimuthal angle $\phi$. Mathematically, this is $R_\phi(\theta) = \exp\left(-i \frac{\theta}{2} (\cos\phi X + \sin\phi Y)\right)$.
    \item \textbf{Arbitrary-Angle ZZ:} \texttt{zz\_phase(q1, q2, $\theta$)} applies the two-qubit entangling gate $ZZ(\theta) = \exp(-i \frac{\theta}{2} Z \otimes Z)$.
\end{itemize}

The following listings provide the compiled instructions for the baseline physical implementation (Listing~\ref{lst:physical_circuit}) and the components of the logically encoded implementations utilizing the Iceberg QED code (Listings~\ref{lst:logical_circuit_0}, ~\ref{lst:logical_circuit_1}, ~\ref{lst:logical_circuit_3}, and \ref{lst:logical_circuit_2}), illustrating how mid-circuit syndrome measurements are interleaved within the computation.

\begin{lstlisting}[language=Python, caption={Compiled physical circuit for the four-state step operator on 4 bare physical qubits. We use the short-hand notation \texttt{rx(q, $theta$)} for \texttt{phased\_x(q, $\theta$, $\phi$)}.}, label={lst:physical_circuit}]
@guppy
def execute_circuit(
        q0: qubit, q1: qubit, q2: qubit, q3: qubit
) -> None:
    rz(q0, 3.50 * pi)
    rz(q1, 3.50 * pi)
    rz(q2, 3.50 * pi)
    rz(q3, 0.50 * pi)
    rx(q0, 1.00 * pi)
    rx(q1, 1.00 * pi)
    rx(q2, 0.25 * pi)
    rx(q3, 0.50 * pi)
    rz(q2, 0.50 * pi)
    rz(q3, 0.50 * pi)
    rz(q2, 3.50 * pi)
    rz(q3, 0.50 * pi)
    rx(q2, 2.50 * pi)
    zz_phase(q3, q2, 0.50 * pi)
    rz(q2, 3.50 * pi)
    rz(q3, 3.50 * pi)
    rx(q2, 3.50 * pi)
    rx(q3, 1.00 * pi)
    rz(q2, 0.50 * pi)
    rz(q2, 3.50 * pi)
    rx(q2, 3.75 * pi)
    rz(q2, 0.50 * pi)
    rz(q2, 0.50 * pi)
    rx(q2, 0.50 * pi)
    rz(q2, 0.50 * pi)
    rz(q2, 1.50 * pi)
    zz_phase(q3, q2, 0.25 * pi)
    rz(q2, 0.75 * pi)
    rz(q3, 1.50 * pi)
    rz(q2, 1.50 * pi)
    rx(q3, 1.00 * pi)
    zz_phase(q0, q2, 0.50 * pi)
    rz(q3, 0.50 * pi)
    rz(q0, 3.50 * pi)
    rz(q2, 1.00 * pi)
    rx(q3, 0.50 * pi)
    rx(q0, 1.00 * pi)
    rz(q2, 1.50 * pi)
    rz(q3, 0.50 * pi)
    rz(q3, 1.50 * pi)
    zz_phase(q1, q3, 0.50 * pi)
    rz(q1, 3.50 * pi)
    rz(q3, 1.00 * pi)
    rx(q1, 1.00 * pi)
    rz(q3, 0.50 * pi)
    rz(q1, 3.50 * pi)
    rx(q3, 0.50 * pi)
    rx(q1, 1.00 * pi)
    rz(q3, 0.50 * pi)
    zz_phase(q1, q2, 0.25 * pi)
    rz(q3, 1.00 * pi)
    rz(q1, 1.50 * pi)
    rz(q2, 0.75 * pi)
    rx(q1, 1.00 * pi)
    rz(q2, 1.00 * pi)
    zz_phase(q3, q2, 0.25 * pi)
    rz(q2, 0.75 * pi)
    rz(q3, 1.00 * pi)
    rz(q2, 0.50 * pi)
    rx(q2, 0.50 * pi)
    rz(q2, 0.50 * pi)

@guppy
def main() -> None:
    """
    Main entry point: Allocates 4 qubits, runs the circuit,
    and returns their classical measurement results.
    """
    q0, q1, q2, q3 = qubit(), qubit(), qubit(), qubit()

    execute_circuit(q0, q1, q2, q3)

    result("x1", measure(q0))
    result("x0", measure(q1))
    result("y1", measure(q2))
    result("y0", measure(q3))
\end{lstlisting}

\begin{lstlisting}[language=Python, caption={Iceberg logically encoded circuit implementation showing the construction of the mid-circuit $S_X$ and $S_Z$ syndrome measurements and the initialization of the logical state into the code space.}, label={lst:logical_circuit_0}]
@guppy
def transpiled_H(i: qubit) -> None:
    rz(i, 1. * pi)
    phased_x(i, 0.50 * pi, 0.50 * pi)

@guppy
def transpiled_CNOT(i: qubit, j: qubit) -> None:
    rz(i, 0.50 * pi)
    rz(j, 3.50 * pi)
    phased_x(j, 2.50 * pi, 0.*pi)
    zz_phase(i, j, .5 * pi)
    phased_x(j, 3.50 * pi, 0.50 * pi)

@guppy
def execute_circuit(
        q0: qubit, q1: qubit, q2: qubit,
        q3: qubit, q4: qubit, q5: qubit
) -> None:
    rz(q0, 1.00 * pi)
    zz_phase(q1, q5, 3.50 * pi)
    phased_x(q0, 0.50 * pi, 0.50 * pi)
    rz(q1, 1.00 * pi)
    zz_phase(q2, q5, 3.50 * pi)
    phased_x(q1, 0.50 * pi, 0.50 * pi)
    rz(q2, 1.00 * pi)
    zz_phase(q3, q5, 3.50 * pi)
    zz_phase(q0, q1, 1.00 * pi)
    phased_x(q2, 0.50 * pi, 0.50 * pi)
    rz(q3, 1.00 * pi)
    zz_phase(q4, q5, 0.50 * pi)
    rz(q0, 1.00 * pi)
    rz(q1, 1.00 * pi)
    phased_x(q3, 0.50 * pi, 0.50 * pi)
    rz(q4, 1.00 * pi)
    phased_x(q0, 0.50 * pi, 0.50 * pi)
    phased_x(q1, 0.50 * pi, 0.50 * pi)
    phased_x(q4, 0.50 * pi, 0.50 * pi)
    rz(q0, 1.00 * pi)
    phased_x(q0, 0.50 * pi, 0.50 * pi)
    zz_phase(q0, q2, 1.00 * pi)
    rz(q0, 1.00 * pi)
    rz(q2, 1.00 * pi)
    phased_x(q0, 0.50 * pi, 0.50 * pi)
    phased_x(q2, 0.50 * pi, 0.50 * pi)
    rz(q0, 1.00 * pi)
    phased_x(q0, 0.50 * pi, 0.50 * pi)
    zz_phase(q0, q3, 0.25 * pi)
    rz(q0, 1.00 * pi)
    rz(q3, 1.00 * pi)
    phased_x(q0, 0.50 * pi, 0.50 * pi)
    phased_x(q3, 0.50 * pi, 0.50 * pi)
    rz(q0, 1.00 * pi)
    zz_phase(q3, q5, 0.50 * pi)
    phased_x(q0, 0.50 * pi, 0.50 * pi)
    zz_phase(q0, q4, 0.50 * pi)
    rz(q0, 1.00 * pi)
    rz(q4, 1.00 * pi)
    phased_x(q0, 0.50 * pi, 0.50 * pi)
    phased_x(q4, 0.50 * pi, 0.50 * pi)
    rz(q0, 1.00 * pi)
    zz_phase(q4, q5, 0.50 * pi)
    phased_x(q0, 0.50 * pi, 0.50 * pi)
    zz_phase(q3, q5, 3.50 * pi)
    rz(q3, 1.00 * pi)
    zz_phase(q4, q5, 0.50 * pi)
    phased_x(q3, 0.50 * pi, 0.50 * pi)
    zz_phase(q0, q3, 2.50 * pi)
    rz(q0, 1.00 * pi)
    rz(q3, 1.00 * pi)
    phased_x(q0, 0.50 * pi, 0.50 * pi)
    phased_x(q3, 0.50 * pi, 0.50 * pi)
    rz(q0, 1.00 * pi)
    zz_phase(q4, q3, 0.50 * pi)
    phased_x(q0, 0.50 * pi, 0.50 * pi)
    zz_phase(q3, q5, 3.50 * pi)
    rz(q3, 1.00 * pi)
    zz_phase(q4, q5, 3.50 * pi)
    phased_x(q3, 0.50 * pi, 0.50 * pi)
    rz(q4, 1.00 * pi)
    zz_phase(q0, q3, 3.50 * pi)
    phased_x(q4, 0.50 * pi, 0.50 * pi)
    rz(q0, 1.00 * pi)
    rz(q3, 1.00 * pi)
    phased_x(q0, 0.50 * pi, 0.50 * pi)
    phased_x(q3, 0.50 * pi, 0.50 * pi)
    rz(q0, 1.00 * pi)
    zz_phase(q3, q5, 0.50 * pi)
    phased_x(q0, 0.50 * pi, 0.50 * pi)
    zz_phase(q3, q5, 3.50 * pi)
    zz_phase(q0, q4, 1.00 * pi)
    rz(q3, 1.00 * pi)
    rz(q0, 1.00 * pi)
    phased_x(q3, 0.50 * pi, 0.50 * pi)
    rz(q4, 1.00 * pi)
    phased_x(q0, 0.50 * pi, 0.50 * pi)
    phased_x(q4, 0.50 * pi, 0.50 * pi)
    rz(q0, 1.00 * pi)
    phased_x(q0, 0.50 * pi, 0.50 * pi)
    zz_phase(q0, q3, 3.75 * pi)
    rz(q0, 1.00 * pi)
    rz(q3, 1.00 * pi)
    phased_x(q0, 0.50 * pi, 0.50 * pi)
    phased_x(q3, 0.50 * pi, 0.50 * pi)
    rz(q0, 1.00 * pi)
    zz_phase(q3, q5, 0.50 * pi)
    phased_x(q0, 0.50 * pi, 0.50 * pi)
    zz_phase(q3, q5, 0.50 * pi)
    rz(q3, 1.00 * pi)
    phased_x(q3, 0.50 * pi, 0.50 * pi)
    zz_phase(q0, q3, 0.50 * pi)
    rz(q0, 1.00 * pi)
    rz(q3, 1.00 * pi)
    phased_x(q0, 0.50 * pi, 0.50 * pi)
    phased_x(q3, 0.50 * pi, 0.50 * pi)
    rz(q0, 1.00 * pi)
    zz_phase(q3, q5, 0.50 * pi)
    phased_x(q0, 0.50 * pi, 0.50 * pi)
    zz_phase(q3, q5, 1.50 * pi)
    zz_phase(q4, q3, 0.25 * pi)
    zz_phase(q3, q5, 0.75 * pi)
    zz_phase(q4, q5, 1.50 * pi)
    zz_phase(q3, q5, 1.50 * pi)
    rz(q4, 1.00 * pi)
    zz_phase(q1, q3, 0.50 * pi)
    phased_x(q4, 0.50 * pi, 0.50 * pi)
    zz_phase(q0, q4, 1.00 * pi)
    rz(q0, 1.00 * pi)
    rz(q4, 1.00 * pi)
    phased_x(q0, 0.50 * pi, 0.50 * pi)
    phased_x(q4, 0.50 * pi, 0.50 * pi)
    rz(q0, 1.00 * pi)
    zz_phase(q4, q5, 0.50 * pi)
    phased_x(q0, 0.50 * pi, 0.50 * pi)
    zz_phase(q1, q5, 3.50 * pi)
    rz(q4, 1.00 * pi)
    rz(q1, 1.00 * pi)
    zz_phase(q3, q5, 1.00 * pi)
    phased_x(q4, 0.50 * pi, 0.50 * pi)
    zz_phase(q0, q4, 0.50 * pi)
    phased_x(q1, 0.50 * pi, 0.50 * pi)
    zz_phase(q3, q5, 1.50 * pi)
    rz(q0, 1.00 * pi)
    rz(q4, 1.00 * pi)
    phased_x(q0, 0.50 * pi, 0.50 * pi)
    phased_x(q4, 0.50 * pi, 0.50 * pi)
    rz(q0, 1.00 * pi)
    zz_phase(q4, q5, 0.50 * pi)
    phased_x(q0, 0.50 * pi, 0.50 * pi)
    zz_phase(q4, q5, 1.50 * pi)
    zz_phase(q0, q1, 1.00 * pi)
    zz_phase(q2, q4, 0.50 * pi)
    rz(q0, 1.00 * pi)
    rz(q1, 1.00 * pi)
    zz_phase(q2, q5, 3.50 * pi)
    phased_x(q0, 0.50 * pi, 0.50 * pi)
    phased_x(q1, 0.50 * pi, 0.50 * pi)
    rz(q2, 1.00 * pi)
    zz_phase(q4, q5, 1.00 * pi)
    rz(q0, 1.00 * pi)
    phased_x(q2, 0.50 * pi, 0.50 * pi)
    zz_phase(q4, q5, 0.50 * pi)
    phased_x(q0, 0.50 * pi, 0.50 * pi)
    rz(q4, 1.00 * pi)
    zz_phase(q0, q2, 1.00 * pi)
    phased_x(q4, 0.50 * pi, 0.50 * pi)
    rz(q0, 1.00 * pi)
    rz(q2, 1.00 * pi)
    phased_x(q0, 0.50 * pi, 0.50 * pi)
    phased_x(q2, 0.50 * pi, 0.50 * pi)
    rz(q0, 1.00 * pi)
    zz_phase(q2, q5, 3.50 * pi)
    phased_x(q0, 0.50 * pi, 0.50 * pi)
    rz(q2, 1.00 * pi)
    zz_phase(q0, q4, 0.50 * pi)
    phased_x(q2, 0.50 * pi, 0.50 * pi)
    rz(q0, 1.00 * pi)
    rz(q4, 1.00 * pi)
    phased_x(q0, 0.50 * pi, 0.50 * pi)
    phased_x(q4, 0.50 * pi, 0.50 * pi)
    rz(q0, 1.00 * pi)
    zz_phase(q4, q5, 0.50 * pi)
    phased_x(q0, 0.50 * pi, 0.50 * pi)
    zz_phase(q4, q5, 1.00 * pi)
    zz_phase(q0, q2, 1.00 * pi)
    rz(q0, 1.00 * pi)
    rz(q2, 1.00 * pi)
    phased_x(q0, 0.50 * pi, 0.50 * pi)
    phased_x(q2, 0.50 * pi, 0.50 * pi)
    rz(q0, 1.00 * pi)
    zz_phase(q2, q3, 0.25 * pi)
    phased_x(q0, 0.50 * pi, 0.50 * pi)
    zz_phase(q2, q5, 1.50 * pi)
    rz(q2, 1.00 * pi)
    zz_phase(q3, q5, 0.75 * pi)
    phased_x(q2, 0.50 * pi, 0.50 * pi)
    zz_phase(q3, q5, 1.00 * pi)
    zz_phase(q0, q2, 1.00 * pi)
    zz_phase(q4, q3, 0.25 * pi)
    rz(q0, 1.00 * pi)
    rz(q2, 1.00 * pi)
    zz_phase(q3, q5, 0.75 * pi)
    phased_x(q0, 0.50 * pi, 0.50 * pi)
    phased_x(q2, 0.50 * pi, 0.50 * pi)
    zz_phase(q4, q5, 1.00 * pi)
    rz(q0, 1.00 * pi)
    zz_phase(q3, q5, 0.50 * pi)
    phased_x(q0, 0.50 * pi, 0.50 * pi)
    rz(q3, 1.00 * pi)
    phased_x(q3, 0.50 * pi, 0.50 * pi)
    zz_phase(q0, q3, 0.50 * pi)
    rz(q0, 1.00 * pi)
    rz(q3, 1.00 * pi)
    phased_x(q0, 0.50 * pi, 0.50 * pi)
    phased_x(q3, 0.50 * pi, 0.50 * pi)
    zz_phase(q3, q5, 0.50 * pi)

@guppy
def main() -> None:
    """
    Main entry point: Allocates 6 qubits, runs the circuit,
    and returns their classical measurement results.
    """
    q0, q1, q2, q3, q4, q5 = qubit(), qubit(), qubit(), qubit(), qubit(), qubit()

    rz(q0, 1. * pi)
    phased_x(q0, 0.50 * pi, 0.50 * pi)
    transpiled_CNOT(q0, q1)
    transpiled_CNOT(q1, q2)
    transpiled_CNOT(q2, q3)
    transpiled_CNOT(q3, q4)
    transpiled_CNOT(q4, q5)

    execute_circuit(q0, q1, q2, q3, q4, q5)

    result("q0", measure(q0))
    result("x1", measure(q0))
    result("x0", measure(q1))
    result("y1", measure(q2))
    result("y0", measure(q3))
    result("q5", measure(q5))
\end{lstlisting}

\begin{lstlisting}[language=Python, caption={Iceberg logically encoded circuit implementation showing the construction of the mid-circuit $S_X$ and $S_Z$ syndrome measurements and the initialization of the logical state into the code space.}, label={lst:logical_circuit_1}]
@guppy
def measure_SZ(q0: qubit, q1: qubit, q2: qubit,
        q3: qubit, q4: qubit, q5: qubit
) -> None:
    sz = qubit()
    transpiled_CNOT(q0, sz)
    transpiled_CNOT(q1, sz)
    transpiled_CNOT(q2, sz)
    transpiled_CNOT(q3, sz)
    transpiled_CNOT(q4, sz)
    transpiled_CNOT(q5, sz)

    result("First ZZZZZZ QED", measure(sz))

@guppy
def measure_SX(q0: qubit, q1: qubit, q2: qubit,
        q3: qubit, q4: qubit, q5: qubit
) -> None:
    transpiled_H(q0)
    transpiled_H(q1)
    transpiled_H(q2)
    transpiled_H(q3)
    transpiled_H(q4)
    transpiled_H(q5)

    sx = qubit()
    transpiled_CNOT(q0, sx)
    transpiled_CNOT(q1, sx)
    transpiled_CNOT(q2, sx)
    transpiled_CNOT(q3, sx)
    transpiled_CNOT(q4, sx)
    transpiled_CNOT(q5, sx)
    result("First XXXXXX QED", measure(sx))

    transpiled_H(q0)
    transpiled_H(q1)
    transpiled_H(q2)
    transpiled_H(q3)
    transpiled_H(q4)
    transpiled_H(q5)

@guppy
def execute_circuit(
        q0: qubit, q1: qubit, q2: qubit,
        q3: qubit, q4: qubit, q5: qubit
) -> None:
    rz(q0, 1.00 * pi)
    zz_phase(q1, q5, 3.50 * pi)
    phased_x(q0, 0.50 * pi, 0.50 * pi)
    rz(q1, 1.00 * pi)
    zz_phase(q2, q5, 3.50 * pi)
    phased_x(q1, 0.50 * pi, 0.50 * pi)
    rz(q2, 1.00 * pi)
    zz_phase(q3, q5, 3.50 * pi)
    zz_phase(q0, q1, 1.00 * pi)
    phased_x(q2, 0.50 * pi, 0.50 * pi)
    rz(q3, 1.00 * pi)
    zz_phase(q4, q5, 0.50 * pi)
    rz(q0, 1.00 * pi)
    rz(q1, 1.00 * pi)
    phased_x(q3, 0.50 * pi, 0.50 * pi)
    rz(q4, 1.00 * pi)
    phased_x(q0, 0.50 * pi, 0.50 * pi)
    phased_x(q1, 0.50 * pi, 0.50 * pi)
    phased_x(q4, 0.50 * pi, 0.50 * pi)
    rz(q0, 1.00 * pi)
    phased_x(q0, 0.50 * pi, 0.50 * pi)
    zz_phase(q0, q2, 1.00 * pi)
    rz(q0, 1.00 * pi)
    rz(q2, 1.00 * pi)
    phased_x(q0, 0.50 * pi, 0.50 * pi)
    phased_x(q2, 0.50 * pi, 0.50 * pi)
    rz(q0, 1.00 * pi)
    phased_x(q0, 0.50 * pi, 0.50 * pi)
    zz_phase(q0, q3, 0.25 * pi)
    rz(q0, 1.00 * pi)
    rz(q3, 1.00 * pi)
    phased_x(q0, 0.50 * pi, 0.50 * pi)
    phased_x(q3, 0.50 * pi, 0.50 * pi)
    rz(q0, 1.00 * pi)
    zz_phase(q3, q5, 0.50 * pi)
    phased_x(q0, 0.50 * pi, 0.50 * pi)
    zz_phase(q0, q4, 0.50 * pi)
    rz(q0, 1.00 * pi)
    rz(q4, 1.00 * pi)
    phased_x(q0, 0.50 * pi, 0.50 * pi)
    phased_x(q4, 0.50 * pi, 0.50 * pi)
    rz(q0, 1.00 * pi)
    zz_phase(q4, q5, 0.50 * pi)
    phased_x(q0, 0.50 * pi, 0.50 * pi)
    zz_phase(q3, q5, 3.50 * pi)
    rz(q3, 1.00 * pi)
    zz_phase(q4, q5, 0.50 * pi)
    phased_x(q3, 0.50 * pi, 0.50 * pi)
    zz_phase(q0, q3, 2.50 * pi)
    rz(q0, 1.00 * pi)
    rz(q3, 1.00 * pi)
    phased_x(q0, 0.50 * pi, 0.50 * pi)
    phased_x(q3, 0.50 * pi, 0.50 * pi)
    rz(q0, 1.00 * pi)
    zz_phase(q4, q3, 0.50 * pi)
    phased_x(q0, 0.50 * pi, 0.50 * pi)
    zz_phase(q3, q5, 3.50 * pi)
    rz(q3, 1.00 * pi)
    zz_phase(q4, q5, 3.50 * pi)
    phased_x(q3, 0.50 * pi, 0.50 * pi)
    rz(q4, 1.00 * pi)
    zz_phase(q0, q3, 3.50 * pi)
    phased_x(q4, 0.50 * pi, 0.50 * pi)
    rz(q0, 1.00 * pi)
    rz(q3, 1.00 * pi)
    phased_x(q0, 0.50 * pi, 0.50 * pi)
    phased_x(q3, 0.50 * pi, 0.50 * pi)
    rz(q0, 1.00 * pi)
    zz_phase(q3, q5, 0.50 * pi)
    phased_x(q0, 0.50 * pi, 0.50 * pi)
    zz_phase(q3, q5, 3.50 * pi)
    zz_phase(q0, q4, 1.00 * pi)
    rz(q3, 1.00 * pi)
    rz(q0, 1.00 * pi)
    phased_x(q3, 0.50 * pi, 0.50 * pi)
    rz(q4, 1.00 * pi)
    phased_x(q0, 0.50 * pi, 0.50 * pi)
    phased_x(q4, 0.50 * pi, 0.50 * pi)
    rz(q0, 1.00 * pi)
    phased_x(q0, 0.50 * pi, 0.50 * pi)
    zz_phase(q0, q3, 3.75 * pi)
    rz(q0, 1.00 * pi)
    rz(q3, 1.00 * pi)
    phased_x(q0, 0.50 * pi, 0.50 * pi)
    phased_x(q3, 0.50 * pi, 0.50 * pi)
    ### SX meas
    measure_SX(q0, q1, q2, q3, q4, q5)
    ###
    ### SZ meas
    measure_SZ(q0, q1, q2, q3, q4, q5)
    ###
    rz(q0, 1.00 * pi)
    zz_phase(q3, q5, 0.50 * pi)
    phased_x(q0, 0.50 * pi, 0.50 * pi)
    zz_phase(q3, q5, 0.50 * pi)
    rz(q3, 1.00 * pi)
    phased_x(q3, 0.50 * pi, 0.50 * pi)
    zz_phase(q0, q3, 0.50 * pi)
    rz(q0, 1.00 * pi)
    rz(q3, 1.00 * pi)
    phased_x(q0, 0.50 * pi, 0.50 * pi)
    phased_x(q3, 0.50 * pi, 0.50 * pi)
    rz(q0, 1.00 * pi)
    zz_phase(q3, q5, 0.50 * pi)
    phased_x(q0, 0.50 * pi, 0.50 * pi)
    zz_phase(q3, q5, 1.50 * pi)
    zz_phase(q4, q3, 0.25 * pi)
    zz_phase(q3, q5, 0.75 * pi)
    zz_phase(q4, q5, 1.50 * pi)
    zz_phase(q3, q5, 1.50 * pi)
    rz(q4, 1.00 * pi)
    zz_phase(q1, q3, 0.50 * pi)
    phased_x(q4, 0.50 * pi, 0.50 * pi)
    zz_phase(q0, q4, 1.00 * pi)
    rz(q0, 1.00 * pi)
    rz(q4, 1.00 * pi)
    phased_x(q0, 0.50 * pi, 0.50 * pi)
    phased_x(q4, 0.50 * pi, 0.50 * pi)
    rz(q0, 1.00 * pi)
    zz_phase(q4, q5, 0.50 * pi)
    phased_x(q0, 0.50 * pi, 0.50 * pi)
    zz_phase(q1, q5, 3.50 * pi)
    rz(q4, 1.00 * pi)
    rz(q1, 1.00 * pi)
    zz_phase(q3, q5, 1.00 * pi)
    phased_x(q4, 0.50 * pi, 0.50 * pi)
    zz_phase(q0, q4, 0.50 * pi)
    phased_x(q1, 0.50 * pi, 0.50 * pi)
    zz_phase(q3, q5, 1.50 * pi)
    rz(q0, 1.00 * pi)
    rz(q4, 1.00 * pi)
    phased_x(q0, 0.50 * pi, 0.50 * pi)
    phased_x(q4, 0.50 * pi, 0.50 * pi)
    rz(q0, 1.00 * pi)
    zz_phase(q4, q5, 0.50 * pi)
    phased_x(q0, 0.50 * pi, 0.50 * pi)
    zz_phase(q4, q5, 1.50 * pi)
    zz_phase(q0, q1, 1.00 * pi)
    zz_phase(q2, q4, 0.50 * pi)
    rz(q0, 1.00 * pi)
    rz(q1, 1.00 * pi)
    zz_phase(q2, q5, 3.50 * pi)
    phased_x(q0, 0.50 * pi, 0.50 * pi)
    phased_x(q1, 0.50 * pi, 0.50 * pi)
    rz(q2, 1.00 * pi)
    zz_phase(q4, q5, 1.00 * pi)
    rz(q0, 1.00 * pi)
    phased_x(q2, 0.50 * pi, 0.50 * pi)
    zz_phase(q4, q5, 0.50 * pi)
    phased_x(q0, 0.50 * pi, 0.50 * pi)
    rz(q4, 1.00 * pi)
    zz_phase(q0, q2, 1.00 * pi)
    phased_x(q4, 0.50 * pi, 0.50 * pi)
    rz(q0, 1.00 * pi)
    rz(q2, 1.00 * pi)
    phased_x(q0, 0.50 * pi, 0.50 * pi)
    phased_x(q2, 0.50 * pi, 0.50 * pi)
    rz(q0, 1.00 * pi)
    zz_phase(q2, q5, 3.50 * pi)
    phased_x(q0, 0.50 * pi, 0.50 * pi)
    rz(q2, 1.00 * pi)
    zz_phase(q0, q4, 0.50 * pi)
    phased_x(q2, 0.50 * pi, 0.50 * pi)
    rz(q0, 1.00 * pi)
    rz(q4, 1.00 * pi)
    phased_x(q0, 0.50 * pi, 0.50 * pi)
    phased_x(q4, 0.50 * pi, 0.50 * pi)
    rz(q0, 1.00 * pi)
    zz_phase(q4, q5, 0.50 * pi)
    phased_x(q0, 0.50 * pi, 0.50 * pi)
    zz_phase(q4, q5, 1.00 * pi)
    zz_phase(q0, q2, 1.00 * pi)
    rz(q0, 1.00 * pi)
    rz(q2, 1.00 * pi)
    phased_x(q0, 0.50 * pi, 0.50 * pi)
    phased_x(q2, 0.50 * pi, 0.50 * pi)
    rz(q0, 1.00 * pi)
    zz_phase(q2, q3, 0.25 * pi)
    phased_x(q0, 0.50 * pi, 0.50 * pi)
    zz_phase(q2, q5, 1.50 * pi)
    rz(q2, 1.00 * pi)
    zz_phase(q3, q5, 0.75 * pi)
    phased_x(q2, 0.50 * pi, 0.50 * pi)
    zz_phase(q3, q5, 1.00 * pi)
    zz_phase(q0, q2, 1.00 * pi)
    zz_phase(q4, q3, 0.25 * pi)
    rz(q0, 1.00 * pi)
    rz(q2, 1.00 * pi)
    zz_phase(q3, q5, 0.75 * pi)
    phased_x(q0, 0.50 * pi, 0.50 * pi)
    phased_x(q2, 0.50 * pi, 0.50 * pi)
    zz_phase(q4, q5, 1.00 * pi)
    rz(q0, 1.00 * pi)
    zz_phase(q3, q5, 0.50 * pi)
    phased_x(q0, 0.50 * pi, 0.50 * pi)
    rz(q3, 1.00 * pi)
    phased_x(q3, 0.50 * pi, 0.50 * pi)
    zz_phase(q0, q3, 0.50 * pi)
    rz(q0, 1.00 * pi)
    rz(q3, 1.00 * pi)
    phased_x(q0, 0.50 * pi, 0.50 * pi)
    phased_x(q3, 0.50 * pi, 0.50 * pi)
    zz_phase(q3, q5, 0.50 * pi)

@guppy
def main() -> None:
    """
    Main entry point: Allocates 6 qubits, runs the circuit,
    and returns their classical measurement results.
    """
    q0, q1, q2, q3, q4, q5 = qubit(), qubit(), qubit(), qubit(), qubit(), qubit()

    transpiled_H(q0)
    transpiled_CNOT(q0, q1)
    transpiled_CNOT(q1, q2)
    transpiled_CNOT(q2, q3)
    transpiled_CNOT(q3, q4)
    transpiled_CNOT(q4, q5)

    execute_circuit(q0, q1, q2, q3, q4, q5)

    result("qt", measure(q0))
    result("x1", measure(q1))
    result("x0", measure(q2))
    result("y1", measure(q3))
    result("y0", measure(q4))
    result("qb", measure(q5))
\end{lstlisting}

\begin{lstlisting}[language=Python, caption={Continuation of the encoded circuit demonstrating the sequential interleaving of mid-circuit operations for up to 5 rounds of active QED.}, label={lst:logical_circuit_3}]
@guppy
def execute_circuit(
        q0: qubit, q1: qubit, q2: qubit,
        q3: qubit, q4: qubit, q5: qubit
) -> None:
    ### SX meas
    measure_SX(q0, q1, q2, q3, q4, q5)
    ###
    ### SZ meas
    measure_SZ(q0, q1, q2, q3, q4, q5)
    ###
    rz(q0, 1.00 * pi)
    zz_phase(q1, q5, 3.50 * pi)
    phased_x(q0, 0.50 * pi, 0.50 * pi)
    rz(q1, 1.00 * pi)
    zz_phase(q2, q5, 3.50 * pi)
    phased_x(q1, 0.50 * pi, 0.50 * pi)
    rz(q2, 1.00 * pi)
    zz_phase(q3, q5, 3.50 * pi)
    zz_phase(q0, q1, 1.00 * pi)
    phased_x(q2, 0.50 * pi, 0.50 * pi)
    rz(q3, 1.00 * pi)
    zz_phase(q4, q5, 0.50 * pi)
    rz(q0, 1.00 * pi)
    rz(q1, 1.00 * pi)
    phased_x(q3, 0.50 * pi, 0.50 * pi)
    rz(q4, 1.00 * pi)
    phased_x(q0, 0.50 * pi, 0.50 * pi)
    phased_x(q1, 0.50 * pi, 0.50 * pi)
    phased_x(q4, 0.50 * pi, 0.50 * pi)
    rz(q0, 1.00 * pi)
    phased_x(q0, 0.50 * pi, 0.50 * pi)
    zz_phase(q0, q2, 1.00 * pi)
    rz(q0, 1.00 * pi)
    rz(q2, 1.00 * pi)
    phased_x(q0, 0.50 * pi, 0.50 * pi)
    phased_x(q2, 0.50 * pi, 0.50 * pi)
    rz(q0, 1.00 * pi)
    phased_x(q0, 0.50 * pi, 0.50 * pi)
    zz_phase(q0, q3, 0.25 * pi)
    rz(q0, 1.00 * pi)
    rz(q3, 1.00 * pi)
    phased_x(q0, 0.50 * pi, 0.50 * pi)
    phased_x(q3, 0.50 * pi, 0.50 * pi)
    rz(q0, 1.00 * pi)
    zz_phase(q3, q5, 0.50 * pi)
    phased_x(q0, 0.50 * pi, 0.50 * pi)
    zz_phase(q0, q4, 0.50 * pi)
    rz(q0, 1.00 * pi)
    rz(q4, 1.00 * pi)
    phased_x(q0, 0.50 * pi, 0.50 * pi)
    phased_x(q4, 0.50 * pi, 0.50 * pi)
    rz(q0, 1.00 * pi)
    zz_phase(q4, q5, 0.50 * pi)
    phased_x(q0, 0.50 * pi, 0.50 * pi)
    zz_phase(q3, q5, 3.50 * pi)
    rz(q3, 1.00 * pi)
    zz_phase(q4, q5, 0.50 * pi)
    phased_x(q3, 0.50 * pi, 0.50 * pi)
    zz_phase(q0, q3, 2.50 * pi)
    rz(q0, 1.00 * pi)
    rz(q3, 1.00 * pi)
    phased_x(q0, 0.50 * pi, 0.50 * pi)
    phased_x(q3, 0.50 * pi, 0.50 * pi)
    rz(q0, 1.00 * pi)
    zz_phase(q4, q3, 0.50 * pi)
    phased_x(q0, 0.50 * pi, 0.50 * pi)
    zz_phase(q3, q5, 3.50 * pi)
    rz(q3, 1.00 * pi)
    zz_phase(q4, q5, 3.50 * pi)
    phased_x(q3, 0.50 * pi, 0.50 * pi)
    rz(q4, 1.00 * pi)
    zz_phase(q0, q3, 3.50 * pi)
    phased_x(q4, 0.50 * pi, 0.50 * pi)
    rz(q0, 1.00 * pi)
    rz(q3, 1.00 * pi)
    phased_x(q0, 0.50 * pi, 0.50 * pi)
    phased_x(q3, 0.50 * pi, 0.50 * pi)
    rz(q0, 1.00 * pi)
    zz_phase(q3, q5, 0.50 * pi)
    phased_x(q0, 0.50 * pi, 0.50 * pi)
    zz_phase(q3, q5, 3.50 * pi)
    zz_phase(q0, q4, 1.00 * pi)
    rz(q3, 1.00 * pi)
    rz(q0, 1.00 * pi)
    phased_x(q3, 0.50 * pi, 0.50 * pi)
    rz(q4, 1.00 * pi)
    phased_x(q0, 0.50 * pi, 0.50 * pi)
    phased_x(q4, 0.50 * pi, 0.50 * pi)
    rz(q0, 1.00 * pi)
    phased_x(q0, 0.50 * pi, 0.50 * pi)
    zz_phase(q0, q3, 3.75 * pi)
    rz(q0, 1.00 * pi)
    rz(q3, 1.00 * pi)
    phased_x(q0, 0.50 * pi, 0.50 * pi)
    phased_x(q3, 0.50 * pi, 0.50 * pi)
    ### SX meas
    measure_SX_2(q0, q1, q2, q3, q4, q5)
    ###
    ### SZ meas
    measure_SZ_2(q0, q1, q2, q3, q4, q5)
    ###
    rz(q0, 1.00 * pi)
    zz_phase(q3, q5, 0.50 * pi)
    phased_x(q0, 0.50 * pi, 0.50 * pi)
    zz_phase(q3, q5, 0.50 * pi)
    rz(q3, 1.00 * pi)
    phased_x(q3, 0.50 * pi, 0.50 * pi)
    zz_phase(q0, q3, 0.50 * pi)
    rz(q0, 1.00 * pi)
    rz(q3, 1.00 * pi)
    phased_x(q0, 0.50 * pi, 0.50 * pi)
    phased_x(q3, 0.50 * pi, 0.50 * pi)
    rz(q0, 1.00 * pi)
    zz_phase(q3, q5, 0.50 * pi)
    phased_x(q0, 0.50 * pi, 0.50 * pi)
    zz_phase(q3, q5, 1.50 * pi)
    zz_phase(q4, q3, 0.25 * pi)
    zz_phase(q3, q5, 0.75 * pi)
    zz_phase(q4, q5, 1.50 * pi)
    zz_phase(q3, q5, 1.50 * pi)
    rz(q4, 1.00 * pi)
    zz_phase(q1, q3, 0.50 * pi)
    phased_x(q4, 0.50 * pi, 0.50 * pi)
    zz_phase(q0, q4, 1.00 * pi)
    rz(q0, 1.00 * pi)
    rz(q4, 1.00 * pi)
    phased_x(q0, 0.50 * pi, 0.50 * pi)
    phased_x(q4, 0.50 * pi, 0.50 * pi)
    rz(q0, 1.00 * pi)
    zz_phase(q4, q5, 0.50 * pi)
    phased_x(q0, 0.50 * pi, 0.50 * pi)
    zz_phase(q1, q5, 3.50 * pi)
    rz(q4, 1.00 * pi)
    rz(q1, 1.00 * pi)
    zz_phase(q3, q5, 1.00 * pi)
    phased_x(q4, 0.50 * pi, 0.50 * pi)
    zz_phase(q0, q4, 0.50 * pi)
    phased_x(q1, 0.50 * pi, 0.50 * pi)
    zz_phase(q3, q5, 1.50 * pi)
    rz(q0, 1.00 * pi)
    rz(q4, 1.00 * pi)
    phased_x(q0, 0.50 * pi, 0.50 * pi)
    phased_x(q4, 0.50 * pi, 0.50 * pi)
    rz(q0, 1.00 * pi)
    zz_phase(q4, q5, 0.50 * pi)
    phased_x(q0, 0.50 * pi, 0.50 * pi)
    zz_phase(q4, q5, 1.50 * pi)
    zz_phase(q0, q1, 1.00 * pi)
    zz_phase(q2, q4, 0.50 * pi)
    rz(q0, 1.00 * pi)
    rz(q1, 1.00 * pi)
    zz_phase(q2, q5, 3.50 * pi)
    phased_x(q0, 0.50 * pi, 0.50 * pi)
    phased_x(q1, 0.50 * pi, 0.50 * pi)
    rz(q2, 1.00 * pi)
    zz_phase(q4, q5, 1.00 * pi)
    rz(q0, 1.00 * pi)
    phased_x(q2, 0.50 * pi, 0.50 * pi)
    zz_phase(q4, q5, 0.50 * pi)
    phased_x(q0, 0.50 * pi, 0.50 * pi)
    rz(q4, 1.00 * pi)
    zz_phase(q0, q2, 1.00 * pi)
    phased_x(q4, 0.50 * pi, 0.50 * pi)
    rz(q0, 1.00 * pi)
    rz(q2, 1.00 * pi)
    phased_x(q0, 0.50 * pi, 0.50 * pi)
    phased_x(q2, 0.50 * pi, 0.50 * pi)
    rz(q0, 1.00 * pi)
    zz_phase(q2, q5, 3.50 * pi)
    phased_x(q0, 0.50 * pi, 0.50 * pi)
    rz(q2, 1.00 * pi)
    zz_phase(q0, q4, 0.50 * pi)
    phased_x(q2, 0.50 * pi, 0.50 * pi)
    rz(q0, 1.00 * pi)
    rz(q4, 1.00 * pi)
    phased_x(q0, 0.50 * pi, 0.50 * pi)
    phased_x(q4, 0.50 * pi, 0.50 * pi)
    rz(q0, 1.00 * pi)
    zz_phase(q4, q5, 0.50 * pi)
    phased_x(q0, 0.50 * pi, 0.50 * pi)
    zz_phase(q4, q5, 1.00 * pi)
    zz_phase(q0, q2, 1.00 * pi)
    rz(q0, 1.00 * pi)
    rz(q2, 1.00 * pi)
    phased_x(q0, 0.50 * pi, 0.50 * pi)
    phased_x(q2, 0.50 * pi, 0.50 * pi)
    ### SX meas
    measure_SX_3(q0, q1, q2, q3, q4, q5)
    ###
    ### SZ meas
    measure_SZ_3(q0, q1, q2, q3, q4, q5)
    ###
    rz(q0, 1.00 * pi)
    zz_phase(q2, q3, 0.25 * pi)
    phased_x(q0, 0.50 * pi, 0.50 * pi)
    zz_phase(q2, q5, 1.50 * pi)
    rz(q2, 1.00 * pi)
    zz_phase(q3, q5, 0.75 * pi)
    phased_x(q2, 0.50 * pi, 0.50 * pi)
    zz_phase(q3, q5, 1.00 * pi)
    zz_phase(q0, q2, 1.00 * pi)
    zz_phase(q4, q3, 0.25 * pi)
    rz(q0, 1.00 * pi)
    rz(q2, 1.00 * pi)
    zz_phase(q3, q5, 0.75 * pi)
    phased_x(q0, 0.50 * pi, 0.50 * pi)
    phased_x(q2, 0.50 * pi, 0.50 * pi)
    zz_phase(q4, q5, 1.00 * pi)
    rz(q0, 1.00 * pi)
    zz_phase(q3, q5, 0.50 * pi)
    phased_x(q0, 0.50 * pi, 0.50 * pi)
    rz(q3, 1.00 * pi)
    phased_x(q3, 0.50 * pi, 0.50 * pi)
    zz_phase(q0, q3, 0.50 * pi)
    rz(q0, 1.00 * pi)
    rz(q3, 1.00 * pi)
    phased_x(q0, 0.50 * pi, 0.50 * pi)
    phased_x(q3, 0.50 * pi, 0.50 * pi)
    zz_phase(q3, q5, 0.50 * pi)
\end{lstlisting}

\begin{lstlisting}[language=Python, caption={Continuation of the encoded circuit demonstrating the sequential interleaving of mid-circuit operations for up to 5 rounds of active QED.}, label={lst:logical_circuit_2}]
@guppy
def execute_circuit(
        q0: qubit, q1: qubit, q2: qubit,
        q3: qubit, q4: qubit, q5: qubit
) -> None:
    ### SX meas
    measure_SX(q0, q1, q2, q3, q4, q5)
    ###
    ### SZ meas
    measure_SZ(q0, q1, q2, q3, q4, q5)
    ###
    rz(q0, 1.00 * pi)
    zz_phase(q1, q5, 3.50 * pi)
    phased_x(q0, 0.50 * pi, 0.50 * pi)
    rz(q1, 1.00 * pi)
    zz_phase(q2, q5, 3.50 * pi)
    phased_x(q1, 0.50 * pi, 0.50 * pi)
    rz(q2, 1.00 * pi)
    zz_phase(q3, q5, 3.50 * pi)
    zz_phase(q0, q1, 1.00 * pi)
    phased_x(q2, 0.50 * pi, 0.50 * pi)
    rz(q3, 1.00 * pi)
    zz_phase(q4, q5, 0.50 * pi)
    rz(q0, 1.00 * pi)
    rz(q1, 1.00 * pi)
    phased_x(q3, 0.50 * pi, 0.50 * pi)
    rz(q4, 1.00 * pi)
    phased_x(q0, 0.50 * pi, 0.50 * pi)
    phased_x(q1, 0.50 * pi, 0.50 * pi)
    phased_x(q4, 0.50 * pi, 0.50 * pi)
    rz(q0, 1.00 * pi)
    phased_x(q0, 0.50 * pi, 0.50 * pi)
    zz_phase(q0, q2, 1.00 * pi)
    rz(q0, 1.00 * pi)
    rz(q2, 1.00 * pi)
    phased_x(q0, 0.50 * pi, 0.50 * pi)
    phased_x(q2, 0.50 * pi, 0.50 * pi)
    rz(q0, 1.00 * pi)
    phased_x(q0, 0.50 * pi, 0.50 * pi)
    zz_phase(q0, q3, 0.25 * pi)
    rz(q0, 1.00 * pi)
    rz(q3, 1.00 * pi)
    phased_x(q0, 0.50 * pi, 0.50 * pi)
    phased_x(q3, 0.50 * pi, 0.50 * pi)
    rz(q0, 1.00 * pi)
    zz_phase(q3, q5, 0.50 * pi)
    phased_x(q0, 0.50 * pi, 0.50 * pi)
    zz_phase(q0, q4, 0.50 * pi)
    rz(q0, 1.00 * pi)
    rz(q4, 1.00 * pi)
    phased_x(q0, 0.50 * pi, 0.50 * pi)
    phased_x(q4, 0.50 * pi, 0.50 * pi)
    ### SX meas
    measure_SX_2(q0, q1, q2, q3, q4, q5)
    ###
    ### SZ meas
    measure_SZ_2(q0, q1, q2, q3, q4, q5)
    ###
    rz(q0, 1.00 * pi)
    zz_phase(q4, q5, 0.50 * pi)
    phased_x(q0, 0.50 * pi, 0.50 * pi)
    zz_phase(q3, q5, 3.50 * pi)
    rz(q3, 1.00 * pi)
    zz_phase(q4, q5, 0.50 * pi)
    phased_x(q3, 0.50 * pi, 0.50 * pi)
    zz_phase(q0, q3, 2.50 * pi)
    rz(q0, 1.00 * pi)
    rz(q3, 1.00 * pi)
    phased_x(q0, 0.50 * pi, 0.50 * pi)
    phased_x(q3, 0.50 * pi, 0.50 * pi)
    rz(q0, 1.00 * pi)
    zz_phase(q4, q3, 0.50 * pi)
    phased_x(q0, 0.50 * pi, 0.50 * pi)
    zz_phase(q3, q5, 3.50 * pi)
    rz(q3, 1.00 * pi)
    zz_phase(q4, q5, 3.50 * pi)
    phased_x(q3, 0.50 * pi, 0.50 * pi)
    rz(q4, 1.00 * pi)
    zz_phase(q0, q3, 3.50 * pi)
    phased_x(q4, 0.50 * pi, 0.50 * pi)
    rz(q0, 1.00 * pi)
    rz(q3, 1.00 * pi)
    phased_x(q0, 0.50 * pi, 0.50 * pi)
    phased_x(q3, 0.50 * pi, 0.50 * pi)
    rz(q0, 1.00 * pi)
    zz_phase(q3, q5, 0.50 * pi)
    phased_x(q0, 0.50 * pi, 0.50 * pi)
    zz_phase(q3, q5, 3.50 * pi)
    zz_phase(q0, q4, 1.00 * pi)
    rz(q3, 1.00 * pi)
    rz(q0, 1.00 * pi)
    phased_x(q3, 0.50 * pi, 0.50 * pi)
    rz(q4, 1.00 * pi)
    phased_x(q0, 0.50 * pi, 0.50 * pi)
    phased_x(q4, 0.50 * pi, 0.50 * pi)
    rz(q0, 1.00 * pi)
    phased_x(q0, 0.50 * pi, 0.50 * pi)
    zz_phase(q0, q3, 3.75 * pi)
    rz(q0, 1.00 * pi)
    rz(q3, 1.00 * pi)
    phased_x(q0, 0.50 * pi, 0.50 * pi)
    phased_x(q3, 0.50 * pi, 0.50 * pi)
    ### SX meas
    measure_SX_3(q0, q1, q2, q3, q4, q5)
    ###
    ### SZ meas
    measure_SZ_3(q0, q1, q2, q3, q4, q5)
    ###
    rz(q0, 1.00 * pi)
    zz_phase(q3, q5, 0.50 * pi)
    phased_x(q0, 0.50 * pi, 0.50 * pi)
    zz_phase(q3, q5, 0.50 * pi)
    rz(q3, 1.00 * pi)
    phased_x(q3, 0.50 * pi, 0.50 * pi)
    zz_phase(q0, q3, 0.50 * pi)
    rz(q0, 1.00 * pi)
    rz(q3, 1.00 * pi)
    phased_x(q0, 0.50 * pi, 0.50 * pi)
    phased_x(q3, 0.50 * pi, 0.50 * pi)
    rz(q0, 1.00 * pi)
    zz_phase(q3, q5, 0.50 * pi)
    phased_x(q0, 0.50 * pi, 0.50 * pi)
    zz_phase(q3, q5, 1.50 * pi)
    zz_phase(q4, q3, 0.25 * pi)
    zz_phase(q3, q5, 0.75 * pi)
    zz_phase(q4, q5, 1.50 * pi)
    zz_phase(q3, q5, 1.50 * pi)
    rz(q4, 1.00 * pi)
    zz_phase(q1, q3, 0.50 * pi)
    phased_x(q4, 0.50 * pi, 0.50 * pi)
    zz_phase(q0, q4, 1.00 * pi)
    rz(q0, 1.00 * pi)
    rz(q4, 1.00 * pi)
    phased_x(q0, 0.50 * pi, 0.50 * pi)
    phased_x(q4, 0.50 * pi, 0.50 * pi)
    ### SX meas
    measure_SX_4(q0, q1, q2, q3, q4, q5)
    ###
    ### SZ meas
    measure_SZ_4(q0, q1, q2, q3, q4, q5)
    ###
    rz(q0, 1.00 * pi)
    zz_phase(q4, q5, 0.50 * pi)
    phased_x(q0, 0.50 * pi, 0.50 * pi)
    zz_phase(q1, q5, 3.50 * pi)
    rz(q4, 1.00 * pi)
    rz(q1, 1.00 * pi)
    zz_phase(q3, q5, 1.00 * pi)
    phased_x(q4, 0.50 * pi, 0.50 * pi)
    zz_phase(q0, q4, 0.50 * pi)
    phased_x(q1, 0.50 * pi, 0.50 * pi)
    zz_phase(q3, q5, 1.50 * pi)
    rz(q0, 1.00 * pi)
    rz(q4, 1.00 * pi)
    phased_x(q0, 0.50 * pi, 0.50 * pi)
    phased_x(q4, 0.50 * pi, 0.50 * pi)
    rz(q0, 1.00 * pi)
    zz_phase(q4, q5, 0.50 * pi)
    phased_x(q0, 0.50 * pi, 0.50 * pi)
    zz_phase(q4, q5, 1.50 * pi)
    zz_phase(q0, q1, 1.00 * pi)
    zz_phase(q2, q4, 0.50 * pi)
    rz(q0, 1.00 * pi)
    rz(q1, 1.00 * pi)
    zz_phase(q2, q5, 3.50 * pi)
    phased_x(q0, 0.50 * pi, 0.50 * pi)
    phased_x(q1, 0.50 * pi, 0.50 * pi)
    ### SX meas
    measure_SX_5(q0, q1, q2, q3, q4, q5)
    ###
    ### SZ meas
    measure_SZ_5(q0, q1, q2, q3, q4, q5)
    ###
    rz(q2, 1.00 * pi)
    zz_phase(q4, q5, 1.00 * pi)
    rz(q0, 1.00 * pi)
    phased_x(q2, 0.50 * pi, 0.50 * pi)
    zz_phase(q4, q5, 0.50 * pi)
    phased_x(q0, 0.50 * pi, 0.50 * pi)
    rz(q4, 1.00 * pi)
    zz_phase(q0, q2, 1.00 * pi)
    phased_x(q4, 0.50 * pi, 0.50 * pi)
    rz(q0, 1.00 * pi)
    rz(q2, 1.00 * pi)
    phased_x(q0, 0.50 * pi, 0.50 * pi)
    phased_x(q2, 0.50 * pi, 0.50 * pi)
    rz(q0, 1.00 * pi)
    zz_phase(q2, q5, 3.50 * pi)
    phased_x(q0, 0.50 * pi, 0.50 * pi)
    rz(q2, 1.00 * pi)
    zz_phase(q0, q4, 0.50 * pi)
    phased_x(q2, 0.50 * pi, 0.50 * pi)
    rz(q0, 1.00 * pi)
    rz(q4, 1.00 * pi)
    phased_x(q0, 0.50 * pi, 0.50 * pi)
    phased_x(q4, 0.50 * pi, 0.50 * pi)
    rz(q0, 1.00 * pi)
    zz_phase(q4, q5, 0.50 * pi)
    phased_x(q0, 0.50 * pi, 0.50 * pi)
    zz_phase(q4, q5, 1.00 * pi)
    zz_phase(q0, q2, 1.00 * pi)
    rz(q0, 1.00 * pi)
    rz(q2, 1.00 * pi)
    phased_x(q0, 0.50 * pi, 0.50 * pi)
    phased_x(q2, 0.50 * pi, 0.50 * pi)
    rz(q0, 1.00 * pi)
    zz_phase(q2, q3, 0.25 * pi)
    phased_x(q0, 0.50 * pi, 0.50 * pi)
    zz_phase(q2, q5, 1.50 * pi)
    rz(q2, 1.00 * pi)
    zz_phase(q3, q5, 0.75 * pi)
    phased_x(q2, 0.50 * pi, 0.50 * pi)
    zz_phase(q3, q5, 1.00 * pi)
    zz_phase(q0, q2, 1.00 * pi)
    zz_phase(q4, q3, 0.25 * pi)
    rz(q0, 1.00 * pi)
    rz(q2, 1.00 * pi)
    zz_phase(q3, q5, 0.75 * pi)
    phased_x(q0, 0.50 * pi, 0.50 * pi)
    phased_x(q2, 0.50 * pi, 0.50 * pi)
    zz_phase(q4, q5, 1.00 * pi)
    rz(q0, 1.00 * pi)
    zz_phase(q3, q5, 0.50 * pi)
    phased_x(q0, 0.50 * pi, 0.50 * pi)
    rz(q3, 1.00 * pi)
    phased_x(q3, 0.50 * pi, 0.50 * pi)
    zz_phase(q0, q3, 0.50 * pi)
    rz(q0, 1.00 * pi)
    rz(q3, 1.00 * pi)
    phased_x(q0, 0.50 * pi, 0.50 * pi)
    phased_x(q3, 0.50 * pi, 0.50 * pi)
    zz_phase(q3, q5, 0.50 * pi)
\end{lstlisting}

\section{Data}\label{app:data}

\begin{table}[h!]
\centering
\begin{tabular}{ccc@{\hspace{1.5cm}}ccc}
\multicolumn{3}{c}{H2-1} & \multicolumn{3}{c}{Helios} \\
\toprule
Outcome & Counts & Probability (\%) & Outcome & Counts & Probability (\%) \\
\midrule
000 & 2586 & 25.86 & 010 & 2600 & 26.00 \\
110 & 2462 & 24.62 & 110 & 2543 & 25.43 \\
100 & 2447 & 24.47 & 100 & 2459 & 24.59 \\
010 & 2376 & 23.76 & 000 & 2321 & 23.21 \\
001 & 42   & 0.42  & 001 & 29   & 0.29  \\
111 & 33   & 0.33  & 101 & 19   & 0.19  \\
011 & 30   & 0.30  & 111 & 16   & 0.16  \\
101 & 24   & 0.24  & 011 & 13   & 0.13  \\
\bottomrule
\end{tabular}
\caption{\centering State preparation using the linear combination of unitaries method on H2-1 (left) and Helios (right). The leftmost bit corresponds denotes the state in $\mathbb S$. The middle bit is $0$ if the state preparation succeeded, $1$ otherwise. The third qubit is the LCU ancilla.}
\label{tab:lcu_state_preparation_combined}
\end{table}

\begin{table*}[h!]
\centering
\begin{tabular}{ccc@{\hspace{1.5cm}}ccc}
\multicolumn{3}{c}{H2-1} & \multicolumn{3}{c}{Helios} \\
\toprule
Outcome & Counts & Probability (\%) & Outcome & Counts & Probability (\%) \\
\midrule
111110 & 140 & 14.00 & 011000 & 134 & 13.40 \\
101110 & 137 & 13.70 & 111110 & 122 & 12.20 \\
011100 & 134 & 13.40 & 101110 & 121 & 12.10 \\
001000 & 119 & 11.90 & 001100 & 118 & 11.80 \\
111010 & 101 & 10.10 & 001000 & 115 & 11.50 \\
101010 & 94  & 9.40  & 111010 & 111 & 11.10 \\
001100 & 93  & 9.30  & 101010 & 99  & 9.90  \\
011000 & 92  & 9.20  & 011100 & 99  & 9.90  \\
010101 & 10  & 1.00  & 000101 & 7   & 0.70  \\
000101 & 9   & 0.90  & 110111 & 6   & 0.60  \\
110011 & 6   & 0.60  & 010001 & 6   & 0.60  \\
000100 & 5   & 0.50  & 110010 & 4   & 0.40  \\
100011 & 5   & 0.50  & 110011 & 4   & 0.40  \\
110111 & 5   & 0.50  & 110110 & 4   & 0.40  \\
010000 & 4   & 0.40  & 000100 & 4   & 0.40  \\
010100 & 4   & 0.40  & 010000 & 4   & 0.40  \\
101111 & 4   & 0.40  & 010100 & 4   & 0.40  \\
110110 & 4   & 0.40  & 100110 & 3   & 0.30  \\
000000 & 3   & 0.30  & 111100 & 3   & 0.30  \\
000001 & 3   & 0.30  & 100010 & 3   & 0.30  \\
001010 & 3   & 0.30  & 001001 & 3   & 0.30  \\
001110 & 3   & 0.30  & 011101 & 2   & 0.20  \\
110010 & 3   & 0.30  & 001110 & 2   & 0.20  \\
010001 & 2   & 0.20  & 000110 & 2   & 0.20  \\
011001 & 2   & 0.20  & 101011 & 2   & 0.20  \\
100111 & 2   & 0.20  & 000001 & 2   & 0.20  \\
101011 & 2   & 0.20  & 100011 & 2   & 0.20  \\
111011 & 2   & 0.20  & 011001 & 1   & 0.10  \\
111111 & 2   & 0.20  & 000000 & 1   & 0.10  \\
001101 & 1   & 0.10  & 010101 & 1   & 0.10  \\
011010 & 1   & 0.10  & 010010 & 1   & 0.10  \\
100110 & 1   & 0.10  & 011010 & 1   & 0.10  \\
101000 & 1   & 0.10  & 111111 & 1   & 0.10  \\
101001 & 1   & 0.10  & 111011 & 1   & 0.10  \\
110100 & 1   & 0.10  & 101111 & 1   & 0.10  \\
111100 & 1   & 0.10  & 011110 & 1   & 0.10  \\
       &     &       & 001101 & 1   & 0.10  \\
       &     &       & 111000 & 1   & 0.10  \\
       &     &       & 110100 & 1   & 0.10  \\
       &     &       & 101100 & 1   & 0.10  \\
       &     &       & 101000 & 1   & 0.10  \\
\bottomrule
\end{tabular}
\caption{\centering Phase estimation measurements on H2-1 (left) and Helios (right). Bit order from left to right: $\mathbb S$, state preparation ancilla, $j_1$, $j_0$, $f$, encoding qubit.}
\label{tab:qpe_combined}
\end{table*}

\begin{table*}[h!]
\centering
\begin{tabular}{ccc@{\hspace{1.5cm}}ccc}
\multicolumn{3}{c}{H2-1} & \multicolumn{3}{c}{Helios} \\
\toprule
Outcome & Count & Probability (\%) & Outcome & Count & Probability (\%) \\
\midrule
111 & 2007 & 20.07 & 000 & 2054 & 20.54 \\
110 & 1779 & 17.79 & 111 & 1813 & 18.13 \\
001 & 1766 & 17.66 & 110 & 1743 & 17.43 \\
000 & 1711 & 17.11 & 001 & 1737 & 17.37 \\
100 & 817  & 8.17  & 011 & 808  & 8.08  \\
101 & 694  & 6.94  & 010 & 662  & 6.62  \\
010 & 625  & 6.25  & 101 & 640  & 6.40  \\
011 & 601  & 6.01  & 100 & 543  & 5.43  \\
\bottomrule
\end{tabular}
\caption{\centering State preparation with Szegedy's method on H2-1 (left) and Helios (right). Bit order: out qubit - in qubit - state preparation ancilla.}
\label{tab:szegedy_state_preparation_combined}
\end{table*}

\begin{table}[h!]
\centering
\begin{tabular}{ccc@{\hspace{1.5cm}}ccc}
\multicolumn{3}{c}{H2-1} & \multicolumn{3}{c}{Helios} \\
\toprule
Outcome & Count & Probability (\%) & Outcome & Count & Probability (\%) \\
\midrule
01000 & 3457 & 34.57 & 00100 & 3527 & 35.27 \\
00100 & 3238 & 32.38 & 01000 & 3286 & 32.86 \\
01010 & 1297 & 12.97 & 00110 & 1330 & 13.30 \\
00110 & 1282 & 12.82 & 01010 & 1262 & 12.62 \\
01100 & 76   & 0.76  & 01100 & 79   & 0.79  \\
10100 & 70   & 0.70  & 00000 & 58   & 0.58  \\
00000 & 54   & 0.54  & 10100 & 49   & 0.49  \\
11000 & 50   & 0.50  & 11000 & 48   & 0.48  \\
00111 & 49   & 0.49  & 00101 & 46   & 0.46  \\
00101 & 48   & 0.48  & 01001 & 43   & 0.43  \\
10110 & 46   & 0.46  & 01011 & 33   & 0.33  \\
01001 & 45   & 0.45  & 10110 & 33   & 0.33  \\
01011 & 43   & 0.43  & 11001 & 26   & 0.26  \\
11010 & 41   & 0.41  & 01110 & 25   & 0.25  \\
10101 & 37   & 0.37  & 11010 & 24   & 0.24  \\
00010 & 33   & 0.33  & 00010 & 23   & 0.23  \\
01110 & 31   & 0.31  & 10101 & 16   & 0.16  \\
11001 & 21   & 0.21  & 00111 & 15   & 0.15  \\
10111 & 20   & 0.20  & 11110 & 14   & 0.14  \\
00001 & 14   & 0.14  & 10111 & 11   & 0.11  \\
11011 & 13   & 0.13  & 11011 & 10   & 0.10  \\
11110 & 9    & 0.09  & 00001 & 9    & 0.09  \\
00011 & 5    & 0.05  & 01101 & 8    & 0.08  \\
01101 & 5    & 0.05  & 00011 & 5    & 0.05  \\
10011 & 5    & 0.05  & 10001 & 5    & 0.05  \\
11101 & 4    & 0.04  & 11100 & 4    & 0.04  \\
01111 & 3    & 0.03  & 10011 & 3    & 0.03  \\
10001 & 2    & 0.02  & 11101 & 2    & 0.02  \\
10000 & 1    & 0.01  & 01111 & 2    & 0.02  \\
10010 & 1    & 0.01  & 11111 & 2    & 0.02  \\
      &      &       & 10010 & 1    & 0.01  \\
      &      &       & 10000 & 1    & 0.01  \\
\bottomrule
\end{tabular}
\caption{\centering State preparation with the controlled-SWAP method on H2-1 (left) and Helios (right). Bit order: state preparation ancilla - in state - out state - coin qubit - $S^c$ ancilla.}
\label{tab:cswap_state_preparation_combined}
\end{table}

\begin{table*}[h!]
\centering
\begin{tabular}{lrr @{\hspace{1cm}} lrr @{\hspace{1cm}} lrr}
\toprule
\multicolumn{3}{c}{H2-1} & \multicolumn{3}{c}{H2-2} & \multicolumn{3}{c}{Helios} \\
\cmidrule(r){1-3} \cmidrule(r){4-6} \cmidrule(r){7-9}
Outcome & Count & Prob. (\%) & Outcome & Count & Prob. (\%) & Outcome & Count & Prob. (\%) \\
\midrule
110010 & 1290 & 12.90 & 101100 & 1311 & 13.11 & 110010 & 1262 & 12.62 \\
110100 & 1285 & 12.85 & 001010 & 1259 & 12.59 & 101010 & 1254 & 12.54 \\
001100 & 1267 & 12.67 & 001100 & 1258 & 12.58 & 010010 & 1251 & 12.51 \\
101010 & 1266 & 12.66 & 010010 & 1243 & 12.43 & 001010 & 1229 & 12.29 \\
101100 & 1233 & 12.33 & 101010 & 1230 & 12.30 & 110100 & 1208 & 12.08 \\
001010 & 1203 & 12.03 & 110010 & 1222 & 12.22 & 001100 & 1197 & 11.97 \\
010100 & 1193 & 11.93 & 010100 & 1216 & 12.16 & 101100 & 1196 & 11.96 \\
010010 & 1171 & 11.71 & 110100 & 1204 & 12.04 & 010100 & 1187 & 11.87 \\
010000 & 9 & 0.09 & 100100 & 8 & 0.08 & 110000 & 52 & 0.52 \\
110000 & 9 & 0.09 & 100010 & 6 & 0.06 & 001000 & 50 & 0.50 \\
111010 & 8 & 0.08 & 000010 & 5 & 0.05 & 010000 & 48 & 0.48 \\
011100 & 7 & 0.07 & 000100 & 5 & 0.05 & 101000 & 45 & 0.45 \\
111100 & 7 & 0.07 & 011100 & 4 & 0.04 & 111100 & 4 & 0.04 \\
000010 & 6 & 0.06 & 101000 & 3 & 0.03 & 101110 & 3 & 0.03 \\
011010 & 5 & 0.05 & 011010 & 3 & 0.03 & 011100 & 3 & 0.03 \\
100100 & 5 & 0.05 & 111100 & 3 & 0.03 & 001101 & 2 & 0.02 \\
000100 & 4 & 0.04 & 010110 & 2 & 0.02 & 100100 & 1 & 0.01 \\
100010 & 4 & 0.04 & 110000 & 2 & 0.02 & 011010 & 1 & 0.01 \\
101000 & 4 & 0.04 & 010000 & 2 & 0.02 & 000010 & 1 & 0.01 \\
010101 & 3 & 0.03 & 110110 & 2 & 0.02 & 001011 & 1 & 0.01 \\
101110 & 3 & 0.03 & 111010 & 2 & 0.02 & 000100 & 1 & 0.01 \\
110110 & 3 & 0.03 & 001000 & 2 & 0.02 & 001110 & 1 & 0.01 \\
001000 & 2 & 0.02 & 001011 & 2 & 0.02 & 101011 & 1 & 0.01 \\
001110 & 2 & 0.02 & 110101 & 2 & 0.02 & 111010 & 1 & 0.01 \\
010110 & 2 & 0.02 & 010011 & 1 & 0.01 & 110110 & 1 & 0.01 \\
101101 & 2 & 0.02 & 110011 & 1 & 0.01 &  &  &  \\
110011 & 2 & 0.02 & 101011 & 1 & 0.01 &  &  &  \\
001011 & 1 & 0.01 & 101110 & 1 & 0.01 &  &  &  \\
001101 & 1 & 0.01 &  &  &  &  &  &  \\
010011 & 1 & 0.01 &  &  &  &  &  &  \\
101011 & 1 & 0.01 &  &  &  &  &  &  \\
110101 & 1 & 0.01 &  &  &  &  &  &  \\
\bottomrule
\end{tabular}
\caption{\centering Measurements of $V\ket0$ on Quantinuum devices H2-1, H2-2, and Helios. The bits denote, from left to right: the SPUE ancilla, outhead, outail, intail, inhead, and coin ancilla.}
\label{tab:state_prep_combined}
\end{table*}

\begin{table}[h!]
\centering
\begin{tabular}{lrr @{\hspace{1cm}} lrr @{\hspace{1cm}} lrr}
\toprule
\multicolumn{3}{c}{H2-1} & \multicolumn{3}{c}{H2-2} & \multicolumn{3}{c}{Helios} \\
\cmidrule(r){1-3} \cmidrule(r){4-6} \cmidrule(r){7-9}
Outcome & Count & Prob. (\%) & Outcome & Count & Prob. (\%) & Outcome & Count & Prob. (\%) \\
\midrule
001010 & 1040 & 10.40 & 110010 & 1114 & 11.14 & 110010 & 1155 & 11.55 \\
101100 & 1028 & 10.28 & 101100 & 1103 & 11.03 & 101100 & 1082 & 10.82 \\
001100 & 1006 & 10.06 & 001010 & 1018 & 10.18 & 001100 & 1076 & 10.76 \\
110100 & 1003 & 10.03 & 001100 & 1013 & 10.13 & 010010 & 1064 & 10.64 \\
010100 & 960  & 9.60  & 010100 & 975  & 9.75  & 101010 & 1026 & 10.26 \\
010010 & 938  & 9.38  & 010010 & 973  & 9.73  & 001010 & 1007 & 10.07 \\
110010 & 934  & 9.34  & 101010 & 901  & 9.01  & 010100 & 956  & 9.56  \\
101010 & 915  & 9.15  & 110100 & 899  & 8.99  & 110100 & 945  & 9.45  \\
110011 & 79   & 0.79  & 110000 & 89   & 0.89  & 110011 & 79   & 0.79  \\
101110 & 76   & 0.76  & 101110 & 73   & 0.73  & 110110 & 77   & 0.77  \\
001000 & 75   & 0.75  & 000101 & 70   & 0.70  & 101110 & 60   & 0.60  \\
010011 & 74   & 0.74  & 010000 & 69   & 0.69  & 101111 & 58   & 0.58  \\
001101 & 71   & 0.71  & 101101 & 62   & 0.62  & 101101 & 54   & 0.54  \\
101011 & 67   & 0.67  & 010011 & 62   & 0.62  & 111101 & 49   & 0.49  \\
001001 & 64   & 0.64  & 111011 & 61   & 0.61  & 001110 & 48   & 0.48  \\
110000 & 64   & 0.64  & 100101 & 63   & 0.63  & 101000 & 46   & 0.46  \\
101001 & 63   & 0.63  & 001000 & 57   & 0.57  & 001101 & 44   & 0.44  \\
101101 & 63   & 0.63  & 101000 & 56   & 0.56  & 110000 & 43   & 0.43  \\
010111 & 59   & 0.59  & 010110 & 55   & 0.55  & 110001 & 42   & 0.42  \\
110111 & 58   & 0.58  & 110011 & 55   & 0.55  & 011010 & 41   & 0.41  \\
\bottomrule
\end{tabular}
\caption{\centering Measurements of $\mathcal{W}V\ket0$ on Quantinuum devices H2-1, H2-2, and Helios. The bits denote, from left to right: the SPUE ancilla, outhead, outail, intail, inhead, and coin ancilla.}
\label{tab:combined_full_circuit}
\end{table}

\begin{table*}[h!]
\centering
\begin{tabular}{lrr @{\hspace{0.5cm}} lrr @{\hspace{0.5cm}} lrr @{\hspace{0.5cm}} lrr}
\toprule
Outcome & Count & (\%) & Outcome & Count & (\%) & Outcome & Count & (\%) & Outcome & Count & (\%) \\
\midrule
0000 & 2463 & 24.63 & 0001 & 2356 & 23.56 & 1110 & 673 & 6.73 & 1101 & 671 & 6.71 \\
1111 & 657  & 6.57  & 1100 & 655  & 6.55  & 0110 & 639 & 6.39 & 0101 & 617 & 6.17 \\
0111 & 609  & 6.09  & 0100 & 588  & 5.88  & 0011 & 18  & 0.18 & 1011 & 17  & 0.17 \\
1010 & 14   & 0.14  & 0010 & 11   & 0.11  & 1001 & 9   & 0.09 & 1000 & 3   & 0.03 \\
\bottomrule
\end{tabular}
\caption{\centering Exhaustive measurement statistics of $O^\dag SO\ket{0000}$ for the $N=2$ lazy simple random walk (state space size $n=4$), executed on Quantinuum H2 devices ($10^4$ shots). The bitstring order is $x_{N-1} \dots x_0y_{N-1} \dots y_0$, corresponding to the binary decompositions $y=\sum_{i=0}^{N-1} y_i2^i$ and $x=\sum_{i=0}^{N-1} x_i2^i$.}
\label{tab:srw_N2_full}
\end{table*}

\begin{table*}[h!]
\centering
\scriptsize
\setlength{\tabcolsep}{4pt}
\begin{tabular}{lrr @{\hspace{0.4cm}} lrr @{\hspace{0.4cm}} lrr @{\hspace{0.4cm}} lrr}
\toprule
Outcome & Count & (\%) & Outcome & Count & (\%) & Outcome & Count & (\%) & Outcome & Count & (\%) \\
\midrule
000001 & 2464 & 24.64 & 000000 & 2460 & 24.60 & 111010 & 651 & 6.51 & 111000 & 626 & 6.26 \\
111001 & 622 & 6.22 & 111011 & 602 & 6.02 & 001010 & 595 & 5.95 & 001001 & 594 & 5.94 \\
001011 & 585 & 5.85 & 001000 & 547 & 5.47 & 000100 & 19 & 0.19 & 000010 & 16 & 0.16 \\
000011 & 16 & 0.16 & 000101 & 15 & 0.15 & 111101 & 14 & 0.14 & 100101 & 13 & 0.13 \\
110111 & 10 & 0.10 & 111111 & 10 & 0.10 & 101101 & 9 & 0.09 & 000111 & 8 & 0.08 \\
000110 & 7 & 0.07 & 100100 & 7 & 0.07 & 111100 & 7 & 0.07 & 111110 & 7 & 0.07 \\
011111 & 6 & 0.06 & 010010 & 5 & 0.05 & 010011 & 5 & 0.05 & 011010 & 5 & 0.05 \\
011100 & 5 & 0.05 & 011101 & 5 & 0.05 & 100001 & 5 & 0.05 & 101100 & 5 & 0.05 \\
101110 & 5 & 0.05 & 101111 & 5 & 0.05 & 001110 & 4 & 0.04 & 001111 & 4 & 0.04 \\
011001 & 4 & 0.04 & 001100 & 3 & 0.03 & 010001 & 3 & 0.03 & 011011 & 3 & 0.03 \\
011110 & 3 & 0.03 & 110000 & 3 & 0.03 & 110110 & 3 & 0.03 & 001101 & 2 & 0.02 \\
100000 & 2 & 0.02 & 110010 & 2 & 0.02 & 110011 & 2 & 0.02 & 110101 & 2 & 0.02 \\
010000 & 1 & 0.01 & 011000 & 1 & 0.01 & 101000 & 1 & 0.01 & 101011 & 1 & 0.01 \\
110001 & 1 & 0.01 & & & & & & & & & \\
\bottomrule
\end{tabular}
\caption{\centering Exhaustive measurement statistics of $O^\dag SO\ket{0000}$ for the $N=3$ lazy simple random walk (state space size $n=8$), executed on Quantinuum H2 devices ($10^4$ shots). The bitstring order is $x_{N-1} \dots x_0y_{N-1} \dots y_0$, corresponding to the binary decompositions $y=\sum_{i=0}^{N-1} y_i2^i$ and $x=\sum_{i=0}^{N-1} x_i2^i$.}
\label{tab:srw_N3_full}
\end{table*}

\begin{table*}[h!]
\centering
\scriptsize
\setlength{\tabcolsep}{3.5pt}
\begin{tabular}{lrr @{\hspace{0.4cm}} lrr @{\hspace{0.4cm}} lrr @{\hspace{0.4cm}} lrr}
\toprule
Outcome & Count & (\%) & Outcome & Count & (\%) & Outcome & Count & (\%) & Outcome & Count & (\%) \\
\midrule
00000001 & 2498 & 24.98 & 00000000 & 2298 & 22.98 & 00010000 & 674 & 6.74 & 00010010 & 626 & 6.26 \\
11110001 & 614 & 6.14 & 11110010 & 604 & 6.04 & 00010011 & 587 & 5.87 & 00010001 & 579 & 5.79 \\
11110000 & 548 & 5.48 & 11110011 & 527 & 5.27 & 00001001 & 17 & 0.17 & 00001000 & 14 & 0.14 \\
10001001 & 12 & 0.12 & 00000011 & 11 & 0.11 & 00001100 & 10 & 0.10 & 00100001 & 9 & 0.09 \\
11111011 & 9 & 0.09 & 00000010 & 8 & 0.08 & 00100010 & 8 & 0.08 & 00000100 & 7 & 0.07 \\
00001101 & 7 & 0.07 & 00010110 & 7 & 0.07 & 01000100 & 7 & 0.07 & 10000001 & 7 & 0.07 \\
10001000 & 7 & 0.07 & 11010011 & 7 & 0.07 & 11110111 & 7 & 0.07 & 00011000 & 6 & 0.06 \\
01011100 & 6 & 0.06 & 10000000 & 6 & 0.06 & 10110001 & 6 & 0.06 & 11001101 & 6 & 0.06 \\
11010111 & 6 & 0.06 & 00000101 & 5 & 0.05 & 00011001 & 5 & 0.05 & 00100000 & 5 & 0.05 \\
00110111 & 5 & 0.05 & 10110010 & 5 & 0.05 & 11001100 & 5 & 0.05 & 11010010 & 5 & 0.05 \\
11100110 & 5 & 0.05 & 00000111 & 4 & 0.04 & 00011011 & 4 & 0.04 & 00110011 & 4 & 0.04 \\
01000001 & 4 & 0.04 & 01111010 & 4 & 0.04 & 10010001 & 4 & 0.04 & 11010100 & 4 & 0.04 \\
11010101 & 4 & 0.04 & 11010110 & 4 & 0.04 & 11100111 & 4 & 0.04 & 11111111 & 4 & 0.04 \\
00000110 & 3 & 0.03 & 01010001 & 3 & 0.03 & 01010010 & 3 & 0.03 & 01011110 & 3 & 0.03 \\
11010000 & 3 & 0.03 & 11100001 & 3 & 0.03 & 11111110 & 3 & 0.03 & 00001010 & 2 & 0.02 \\
00010100 & 2 & 0.02 & 00011100 & 2 & 0.02 & 00011110 & 2 & 0.02 & 00100111 & 2 & 0.02 \\
00110000 & 2 & 0.02 & 01010000 & 2 & 0.02 & 01010100 & 2 & 0.02 & 01111000 & 2 & 0.02 \\
10010010 & 2 & 0.02 & 10011000 & 2 & 0.02 & 10011001 & 2 & 0.02 & 10011010 & 2 & 0.02 \\
10011011 & 2 & 0.02 & 10111010 & 2 & 0.02 & 10111101 & 2 & 0.02 & 11010001 & 2 & 0.02 \\
11100010 & 2 & 0.02 & 11100011 & 2 & 0.02 & 00010101 & 1 & 0.01 & 00011101 & 1 & 0.01 \\
00100110 & 1 & 0.01 & 00101010 & 1 & 0.01 & 01011101 & 1 & 0.01 & 01101110 & 1 & 0.01 \\
01111001 & 1 & 0.01 & 11111101 & 1 & 0.01 & 00101110 & 1 & 0.01 & 00110001 & 1 & 0.01 \\
00110010 & 1 & 0.01 & 00110101 & 1 & 0.01 & 01000010 & 1 & 0.01 & 01000110 & 1 & 0.01 \\
01001101 & 1 & 0.01 & 01010011 & 1 & 0.01 & 01111111 & 1 & 0.01 & 10001100 & 1 & 0.01 \\
10010000 & 1 & 0.01 & 10100000 & 1 & 0.01 & 10101010 & 1 & 0.01 & 10110101 & 1 & 0.01 \\
10111000 & 1 & 0.01 & 10111001 & 1 & 0.01 & 10111100 & 1 & 0.01 & 10111111 & 1 & 0.01 \\
11000110 & 1 & 0.01 & 11011010 & 1 & 0.01 & 11011011 & 1 & 0.01 & 11011110 & 1 & 0.01 \\
11011111 & 1 & 0.01 & 11100000 & 1 & 0.01 & 11100100 & 1 & 0.01 & 11101011 & 1 & 0.01 \\
11101100 & 1 & 0.01 & 11101110 & 1 & 0.01 & 11110100 & 1 & 0.01 & & & \\
\bottomrule
\end{tabular}
\caption{\centering Exhaustive measurement statistics of $O^\dag SO\ket{0000}$ for the $N=4$ lazy simple random walk (state space size $n=16$), executed on Quantinuum H2 devices ($10^4$ shots). The bitstring order is $x_{N-1} \dots x_0y_{N-1} \dots y_0$, corresponding to the binary decompositions $y=\sum_{i=0}^{N-1} y_i2^i$ and $x=\sum_{i=0}^{N-1} x_i2^i$.}
\label{tab:srw_N4_full}
\end{table*}

\begin{table*}[h!]
\centering
\scriptsize
\setlength{\tabcolsep}{3pt}
\begin{tabular}{lrr @{\hspace{0.4cm}} lrr @{\hspace{0.4cm}} lrr @{\hspace{0.4cm}} lrr}
\toprule
Outcome & Count & (\%) & Outcome & Count & (\%) & Outcome & Count & (\%) & Outcome & Count & (\%) \\
\midrule
0000000000 & 2352 & 23.52 & 0000000001 & 2326 & 23.26 & 0000100000 & 633 & 6.33 & 1111100000 & 611 & 6.11 \\
0000100001 & 605 & 6.05 & 1111100010 & 587 & 5.87 & 0000100010 & 586 & 5.86 & 0000100011 & 576 & 5.76 \\
1111100001 & 569 & 5.69 & 1111100011 & 546 & 5.46 & 0000010001 & 17 & 0.17 & 0000000011 & 14 & 0.14 \\
1000010000 & 14 & 0.14 & 0000001000 & 13 & 0.13 & 0001000000 & 13 & 0.13 & 0000001100 & 12 & 0.12 \\
0000100000 & 12 & 0.12 & 0100001000 & 12 & 0.12 & 0000000010 & 11 & 0.11 & 1000010001 & 11 & 0.11 \\
1111100111 & 10 & 0.10 & 0000000100 & 9 & 0.09 & 0000000111 & 9 & 0.09 & 0100001001 & 9 & 0.09 \\
1111111011 & 9 & 0.09 & 0010000100 & 8 & 0.08 & 0111111000 & 8 & 0.08 & 0000000101 & 7 & 0.07 \\
0000001101 & 7 & 0.07 & 0000011000 & 7 & 0.07 & 0100001111 & 7 & 0.07 & 0000101000 & 6 & 0.06 \\
0000110000 & 6 & 0.06 & 0000110011 & 6 & 0.06 & 0001000001 & 6 & 0.06 & 0010000001 & 6 & 0.06 \\
0100000000 & 6 & 0.06 & 0100000001 & 6 & 0.06 & 1011000101 & 6 & 0.06 & 1101100000 & 6 & 0.06 \\
0000011001 & 5 & 0.05 & 0001100100 & 5 & 0.05 & 0001100110 & 5 & 0.05 & 0010000101 & 5 & 0.05 \\
0010101100 & 5 & 0.05 & 0010101111 & 5 & 0.05 & 0011101010 & 5 & 0.05 & 1000110001 & 5 & 0.05 \\
1000110011 & 5 & 0.05 & 1011001001 & 5 & 0.05 & 1101001010 & 5 & 0.05 & 1110001100 & 5 & 0.05 \\
1110100111 & 5 & 0.05 & 0001000010 & 4 & 0.04 & 0001100111 & 4 & 0.04 & 0010000010 & 4 & 0.04 \\
0010001001 & 4 & 0.04 & 0111110000 & 4 & 0.04 & 0111110010 & 4 & 0.04 & 0111110011 & 4 & 0.04 \\
1000000000 & 4 & 0.04 & 1000000001 & 4 & 0.04 & 1000100000 & 4 & 0.04 & 1000110010 & 4 & 0.04 \\
1011000000 & 4 & 0.04 & 1101010000 & 4 & 0.04 & 1101010101 & 4 & 0.04 & 1101011000 & 4 & 0.04 \\
1101011100 & 4 & 0.04 & 1110100011 & 4 & 0.04 & 1111010001 & 4 & 0.04 & 1111010011 & 4 & 0.04 \\
1111110001 & 4 & 0.04 & 1111111000 & 4 & 0.04 & 1111111001 & 4 & 0.04 & 0000100100 & 3 & 0.03 \\
0000100110 & 3 & 0.03 & 0000101110 & 3 & 0.03 & 0000111011 & 3 & 0.03 & 0001000001 & 3 & 0.03 \\
0001000111 & 3 & 0.03 & 0001100011 & 3 & 0.03 & 0001100101 & 3 & 0.03 & 0010100010 & 3 & 0.03 \\
0010101110 & 3 & 0.03 & 0100010010 & 3 & 0.03 & 0100010011 & 3 & 0.03 & 0100100010 & 3 & 0.03 \\
0100100011 & 3 & 0.03 & 0100101011 & 3 & 0.03 & 0100111010 & 3 & 0.03 & 0100111011 & 3 & 0.03 \\
0111100011 & 3 & 0.03 & 1000100001 & 3 & 0.03 & 1101101100 & 3 & 0.03 & 1110100110 & 3 & 0.03 \\
1111000111 & 3 & 0.03 & 1111100101 & 3 & 0.03 & 1111101011 & 3 & 0.03 & 1111101111 & 3 & 0.03 \\
1111111010 & 3 & 0.03 & 0000111110 & 2 & 0.02 & 0100010001 & 2 & 0.02 & 1011110011 & 2 & 0.02 \\
1110110110 & 2 & 0.02 & 1000001100 & 2 & 0.02 & 1110110011 & 2 & 0.02 & 1110111000 & 2 & 0.02 \\
0101111100 & 2 & 0.02 & 0010011100 & 2 & 0.02 & 1110000010 & 2 & 0.02 & 1000100011 & 2 & 0.02 \\
0011001110 & 2 & 0.02 & 1010110011 & 2 & 0.02 & 1011100010 & 2 & 0.02 & 1011100011 & 2 & 0.02 \\
1101100011 & 2 & 0.02 & 1101100111 & 2 & 0.02 & 1101101111 & 2 & 0.02 & 1111000100 & 2 & 0.02 \\
1111000101 & 2 & 0.02 & 1111001011 & 2 & 0.02 & 1111011001 & 2 & 0.02 & 1111100010 & 2 & 0.02 \\
1111100011 & 2 & 0.02 & 0001101100 & 1 & 0.01 & 0011100011 & 1 & 0.01 & 0100101100 & 1 & 0.01 \\
1011101001 & 1 & 0.01 & 0011100110 & 1 & 0.01 & 1000111011 & 1 & 0.01 & 0000101100 & 1 & 0.01 \\
0011101011 & 1 & 0.01 & 1110100100 & 1 & 0.01 & 1111101001 & 1 & 0.01 & 0000001110 & 1 & 0.01 \\
0000111000 & 1 & 0.01 & 0010100101 & 1 & 0.01 & 0011110001 & 1 & 0.01 & 0100000101 & 1 & 0.01 \\
1010111010 & 1 & 0.01 & 1010111110 & 1 & 0.01 & 1011011110 & 1 & 0.01 & 1100001001 & 1 & 0.01 \\
1100010000 & 1 & 0.01 & 1110101100 & 1 & 0.01 & 1111101010 & 1 & 0.01 & 0000010100 & 1 & 0.01 \\
0000010101 & 1 & 0.01 & 0001000110 & 1 & 0.01 & 0011111000 & 1 & 0.01 & 0100101010 & 1 & 0.01 \\
1001110110 & 1 & 0.01 & 1001110111 & 1 & 0.01 & 1111110101 & 1 & 0.01 & 1111111101 & 1 & 0.01 \\
0001010000 & 1 & 0.01 & 0001100001 & 1 & 0.01 & 0100111000 & 1 & 0.01 & 0100111001 & 1 & 0.01 \\
0111010110 & 1 & 0.01 & 1000110100 & 1 & 0.01 & 1110110111 & 1 & 0.01 & 1110111001 & 1 & 0.01 \\
1111010011 & 1 & 0.01 & 1111101110 & 1 & 0.01 & 0010111001 & 1 & 0.01 & 0011101000 & 1 & 0.01 \\
\bottomrule
\end{tabular}
\caption{\centering Exhaustive measurement statistics of $O^\dag SO\ket{0000}$ for the $N=5$ lazy simple random walk (state space size $n=32$), executed on Quantinuum H2 devices ($10^4$ shots). The bitstring order is $x_{N-1} \dots x_0y_{N-1} \dots y_0$, corresponding to the binary decompositions $y=\sum_{i=0}^{N-1} y_i2^i$ and $x=\sum_{i=0}^{N-1} x_i2^i$.}
\label{tab:srw_N5_full}
\end{table*}

\begin{table*}[h!]
\centering
\scriptsize
\setlength{\tabcolsep}{2pt}
\begin{tabular}{lrr @{\hspace{0.2cm}} lrr @{\hspace{0.2cm}} lrr @{\hspace{0.2cm}} lrr}
\toprule
Outcome & Count & (\%) & Outcome & Count & (\%) & Outcome & Count & (\%) & Outcome & Count & (\%) \\
\midrule
000000000001 & 2325 & 23.25 & 000000000000 & 2311 & 23.11 & 111111000011 & 611 & 6.11 & 111111000000 & 573 & 5.73 \\
000001000010 & 569 & 5.69 & 111111000001 & 564 & 5.64 & 000001000001 & 557 & 5.57 & 000001000000 & 550 & 5.50 \\
000001000011 & 549 & 5.49 & 111111000010 & 538 & 5.38 & 000000000011 & 19 & 0.19 & 000000100001 & 18 & 0.18 \\
111111000111 & 16 & 0.16 & 000000000010 & 15 & 0.15 & 100000100001 & 15 & 0.15 & 000000001010 & 14 & 0.14 \\
100001000000 & 14 & 0.14 & 010000010000 & 13 & 0.13 & 000000000110 & 12 & 0.12 & 000010000010 & 12 & 0.12 \\
100000100000 & 12 & 0.12 & 000000010001 & 11 & 0.11 & 000000100000 & 11 & 0.11 & 100001000001 & 11 & 0.11 \\
000010000000 & 10 & 0.10 & 001000001000 & 10 & 0.10 & 111110000111 & 10 & 0.10 & 000000000101 & 9 & 0.09 \\
000010000011 & 9 & 0.09 & 010000010001 & 9 & 0.09 & 000000010000 & 8 & 0.08 & 001000000000 & 8 & 0.08 \\
011111100000 & 8 & 0.08 & 000000000100 & 7 & 0.07 & 000000001101 & 7 & 0.07 & 000000011000 & 7 & 0.07 \\
000000110000 & 7 & 0.07 & 000000110001 & 7 & 0.07 & 000001010011 & 7 & 0.07 & 000001110010 & 7 & 0.07 \\
000010000001 & 7 & 0.07 & 000000001001 & 7 & 0.07 & 110000110000 & 6 & 0.06 & 111100001100 & 6 & 0.06 \\
111101000000 & 6 & 0.06 & 000101001110 & 5 & 0.05 & 111011000000 & 5 & 0.05 & 111011000010 & 5 & 0.05 \\
111011000011 & 5 & 0.05 & 000101001100 & 4 & 0.04 & 101111000010 & 4 & 0.04 & 110000110001 & 4 & 0.04 \\
111011000001 & 4 & 0.04 & 111011001110 & 4 & 0.04 & 111101000001 & 4 & 0.04 & 111101000011 & 4 & 0.04 \\
111111001101 & 4 & 0.04 & 101111000010 & 4 & 0.04 & 110000110001 & 4 & 0.04 & 111011000001 & 4 & 0.04 \\
111011001110 & 4 & 0.04 & 111101000001 & 4 & 0.04 & 111101000011 & 4 & 0.04 & 111111001101 & 4 & 0.04 \\
000101000101 & 1 & 0.01 & 000101000110 & 1 & 0.01 & 001000001001 & 1 & 0.01 & 001000010100 & 1 & 0.01 \\
101111110010 & 1 & 0.01 & 101111000000 & 3 & 0.03 & 110111011001 & 3 & 0.03 & 111000011001 & 3 & 0.03 \\
111011001101 & 3 & 0.03 & 111101000100 & 3 & 0.03 & 1111101011 & 3 & 0.03 & 111111010000 & 3 & 0.03 \\
111111010011 & 3 & 0.03 & 110000110000 & 6 & 0.06 & 111100001100 & 6 & 0.06 & 111101000000 & 6 & 0.06 \\
000101001110 & 5 & 0.05 & 111011000000 & 5 & 0.05 & 111011000010 & 5 & 0.05 & 111011000011 & 5 & 0.05 \\
000101001100 & 4 & 0.04 & 001111010010 & 2 & 0.02 & 111111000101 & 6 & 0.06 & 111111001010 & 2 & 0.02 \\
111111001011 & 1 & 0.01 & 010001000000 & 1 & 0.01 & 010001000001 & 1 & 0.01 & 010001010000 & 1 & 0.01 \\
010001010011 & 1 & 0.01 & 010001110010 & 1 & 0.01 & 010101110100 & 1 & 0.01 & 011001010011 & 1 & 0.01 \\
011011110100 & 1 & 0.01 & 011100101100 & 1 & 0.01 & 011100101101 & 1 & 0.01 & 011101100100 & 1 & 0.01 \\
011101100110 & 1 & 0.01 & 011111000000 & 1 & 0.01 & 011111010010 & 1 & 0.01 & 011111100001 & 1 & 0.01 \\
011111100011 & 1 & 0.01 & 100000010001 & 1 & 0.01 & 100000111001 & 1 & 0.01 & 100001100000 & 1 & 0.01 \\
100001100011 & 1 & 0.01 & 100001100100 & 1 & 0.01 & 100001110011 & 1 & 0.01 & 100010000001 & 1 & 0.01 \\
100011100111 & 1 & 0.01 & 100101101110 & 1 & 0.01 & 100111101010 & 1 & 0.01 & 101000101000 & 1 & 0.01 \\
101000101001 & 1 & 0.01 & 101001100010 & 1 & 0.01 & 101010100100 & 1 & 0.01 & 101011111100 & 1 & 0.01 \\
101100111100 & 1 & 0.01 & 101101000001 & 1 & 0.01 & 101101000110 & 1 & 0.01 & 110000010000 & 1 & 0.01 \\
110001010010 & 1 & 0.01 & 110001110111 & 1 & 0.01 & 110011001100 & 1 & 0.01 & 110100010100 & 1 & 0.01 \\
110101101010 & 1 & 0.01 & 110111000000 & 1 & 0.01 & 110111000001 & 1 & 0.01 & 110111000011 & 1 & 0.01 \\
110111001010 & 1 & 0.01 & 110111011000 & 1 & 0.01 & 110111011110 & 1 & 0.01 & 110111101010 & 1 & 0.01 \\
111000011110 & 1 & 0.01 & 111001001001 & 1 & 0.01 & 111010001010 & 1 & 0.01 & 111011001111 & 1 & 0.01 \\
111011011001 & 1 & 0.01 & 111011011010 & 1 & 0.01 & 111011011011 & 1 & 0.01 & 111011100001 & 1 & 0.01 \\
111100001101 & 1 & 0.01 & 111101000101 & 1 & 0.01 & 111101001011 & 1 & 0.01 & 111101001100 & 1 & 0.01 \\
111110000010 & 1 & 0.01 & 111110000011 & 1 & 0.01 & 111110001100 & 1 & 0.01 & 111110110010 & 1 & 0.01 \\
111111000100 & 3 & 0.03 & 111111011011 & 1 & 0.01 & 111111010101 & 1 & 0.01 & 001000010001 & 2 & 0.02 \\
010011101011 & 2 & 0.02 & 010011101101 & 2 & 0.02 & 010011110001 & 2 & 0.02 & 011010100010 & 2 & 0.02 \\
100011001101 & 2 & 0.02 & 101010000010 & 2 & 0.02 & 101101011100 & 2 & 0.02 & 110000010010 & 2 & 0.02 \\
110000110011 & 2 & 0.02 & \textit{+ 62 others} & 1 & 0.01 & & & & & & \\
\bottomrule
\end{tabular}
\caption{\centering Exhaustive measurement statistics of $O^\dag SO\ket{0000}$ for the $N=6$ lazy simple random walk (state space size $n=64$), executed on Quantinuum H2 devices ($10^4$ shots). 12 total bits. The bitstring order is $x_{N-1} \dots x_0y_{N-1} \dots y_0$, corresponding to the binary decompositions $y=\sum_{i=0}^{N-1} y_i2^i$ and $x=\sum_{i=0}^{N-1} x_i2^i$.}
\label{tab:srw_N6_full}
\end{table*}

\begin{table*}[h!]
\centering
\begin{tabular}{lrr @{\hspace{1cm}} lrr @{\hspace{1cm}} lrr}
\toprule
Outcome & Count & Prob. (\%) & Outcome & Count & Prob. (\%) & Outcome & Count & Prob. (\%) \\
\midrule
0000 & 490 & 49.00 & 0001 & 236 & 23.60 & 0110 & 3 & 0.30 \\
0011 & 264 & 26.40 & 0010 & 6   & 0.60  & 0101 & 1 & 0.10 \\
\bottomrule
\end{tabular}
\caption{\centering Measurement statistics corresponding to the direct physical sampling of the step operator $O\ket{0000}$ for the four-state random walk ($10^3$ shots). The baseline circuit (0 rounds of error detection) was executed. The bitstring order is $x_1 x_0 y_1 y_0$, where $x$ represents the initial state register (which ideally remains $00$) and $y$ represents the target register undergoing the walk transitions.}
\label{tab:srw_order0}
\end{table*}

\begin{table*}[h!]
\centering
\scriptsize
\begin{tabular}{lrr @{\hspace{0.8cm}} lrr @{\hspace{0.8cm}} lrr}
\toprule
Outcome & Count & Prob. (\%) & Outcome & Count & Prob. (\%) & Outcome & Count & Prob. (\%) \\
\midrule
111111 & 269 & 26.90 & 011011 & 5 & 0.50 & 101100 & 1 & 0.10 \\
000000 & 245 & 24.50 & 011010 & 4 & 0.40 & 000101 & 1 & 0.10 \\
011101 & 121 & 12.10 & 100000 & 3 & 0.30 & 011000 & 1 & 0.10 \\
100010 & 109 & 10.90 & 000001 & 3 & 0.30 & 001000 & 1 & 0.10 \\
111001 & 107 & 10.70 & 000111 & 3 & 0.30 & 001001 & 1 & 0.10 \\
000110 & 106 & 10.60 & 000100 & 2 & 0.20 & 011001 & 1 & 0.10 \\
100100 & 6   & 0.60  & 000010 & 2 & 0.20 & 101001 & 1 & 0.10 \\
100011 & 5   & 0.50  & 100110 & 2 & 0.20 & 011100 & 1 & 0.10 \\
\bottomrule
\end{tabular}
\caption{\centering Exhaustive measurement statistics for the step operator $O\ket{0000}$ utilizing Iceberg QED codes with post-selection only (0 rounds of mid-circuit error detection, $10^3$ total shots). The bitstring order is $q_t \, x_1 \, x_0 \, y_1 \, y_0 \, q_b$. Here, $x$ and $y$ designate the initial and target logical registers, and $q_t$ and $q_b$ represent the top and bottom parity measurement qubits extracted at the end of the circuit.}
\label{tab:srw_order1_postselect}
\end{table*}

\begin{table*}[h!]
\centering
\scriptsize
\begin{tabular}{lrr @{\hspace{0.8cm}} lrr @{\hspace{0.8cm}} lrr}
\toprule
Outcome & Count & Prob. (\%) & Outcome & Count & Prob. (\%) & Outcome & Count & Prob. (\%) \\
\midrule
00000000 & 233 & 23.30 & 01011000 & 2 & 0.20 & 11101111 & 1 & 0.10 \\
11111100 & 228 & 22.80 & 00011001 & 2 & 0.20 & 00011110 & 1 & 0.10 \\
01110100 & 117 & 11.70 & 11100101 & 2 & 0.20 & 01111101 & 1 & 0.10 \\
00011000 & 116 & 11.60 & 10001100 & 2 & 0.20 & 01110010 & 1 & 0.10 \\
10001000 & 114 & 11.40 & 00010000 & 2 & 0.20 & 11111001 & 1 & 0.10 \\
11100100 & 108 & 10.80 & 00000100 & 2 & 0.20 & 01101111 & 1 & 0.10 \\
11111110 & 11  & 1.10  & 10101100 & 1 & 0.10 & 00010011 & 1 & 0.10 \\
01101100 & 4   & 0.40  & 00011101 & 1 & 0.10 & 10001010 & 1 & 0.10 \\
11111101 & 3   & 0.30  & 10011001 & 1 & 0.10 & 00010110 & 1 & 0.10 \\
00000010 & 3   & 0.30  & 11110000 & 1 & 0.10 & 11101101 & 1 & 0.10 \\
11100011 & 3   & 0.30  & 11010001 & 1 & 0.10 & 01011011 & 1 & 0.10 \\
10001101 & 2   & 0.20  & 10010001 & 1 & 0.10 & 10000001 & 1 & 0.10 \\
10101001 & 2   & 0.20  & 00001000 & 1 & 0.10 & 11101110 & 1 & 0.10 \\
01110110 & 2   & 0.20  & 01001000 & 1 & 0.10 & 11011101 & 1 & 0.10 \\
10010111 & 2   & 0.20  & 00010001 & 1 & 0.10 & 11110100 & 1 & 0.10 \\
11100110 & 2   & 0.20  & 11010011 & 1 & 0.10 & 01000000 & 1 & 0.10 \\
00011010 & 2   & 0.20  & 11000100 & 1 & 0.10 & 01001101 & 1 & 0.10 \\
10010000 & 2   & 0.20  & 10011000 & 1 & 0.10 & 10101000 & 1 & 0.10 \\
10000000 & 2   & 0.20  & 01111100 & 1 & 0.10 & 00110111 & 1 & 0.10 \\
\bottomrule
\end{tabular}
\caption{\centering Exhaustive measurement statistics for the step operator $O\ket{0000}$ utilizing Iceberg QED codes with 1 round of mid-circuit error detection ($10^3$ total shots, out of which $942$ successfully passed the QED validation). The bitstring order is $q_t \, x_1 \, x_0 \, y_1 \, y_0 \, q_b \, s_X \, s_Z$. Here, $x$ and $y$ designate the initial and target logical registers, $q_t$ and $q_b$ represent the top and bottom parity qubits, and $s_X$ and $s_Z$ correspond to the $X$ and $Z$ mid-circuit QED syndrome measurement outcomes, respectively.}
\label{tab:srw_order1_qed1}
\end{table*}

\begin{table*}[h!]
\centering
\scriptsize
\begin{tabular}{lrr @{\hspace{0.8cm}} lrr @{\hspace{0.8cm}} lrr}
\toprule
Outcome & Count & Prob. (\%) & Outcome & Count & Prob. (\%) & Outcome & Count & Prob. (\%) \\
\midrule
111111000000 & 234 & 23.40 & 101100000101 & 1 & 0.10 & 011111000101 & 1 & 0.10 \\
000000000000 & 220 & 22.00 & 011100000011 & 1 & 0.10 & 101100110101 & 1 & 0.10 \\
000110000000 & 116 & 11.60 & 100100001110 & 1 & 0.10 & 101001000001 & 1 & 0.10 \\
100010000000 & 107 & 10.70 & 010101000010 & 1 & 0.10 & 100011000110 & 1 & 0.10 \\
011101000000 & 93  & 9.30  & 100100111110 & 1 & 0.10 & 000100000011 & 1 & 0.10 \\
111001000000 & 89  & 8.90  & 000100001111 & 1 & 0.10 & 111111000001 & 1 & 0.10 \\
000000001010 & 6   & 0.60  & 100000000101 & 1 & 0.10 & 110111001101 & 1 & 0.10 \\
000110000010 & 5   & 0.50  & 100101000011 & 1 & 0.10 & 010000001101 & 1 & 0.10 \\
111111000010 & 5   & 0.50  & 100011000011 & 1 & 0.10 & 111101110100 & 1 & 0.10 \\
000000000010 & 4   & 0.40  & 000010000000 & 1 & 0.10 & 000010000001 & 1 & 0.10 \\
100010001010 & 4   & 0.40  & 000111000011 & 1 & 0.10 & 100110000000 & 1 & 0.10 \\
111111001010 & 4   & 0.40  & 011111000011 & 1 & 0.10 & 111111100000 & 1 & 0.10 \\
011101000010 & 3   & 0.30  & 000100010101 & 1 & 0.10 & 100011010101 & 1 & 0.10 \\
111001000010 & 3   & 0.30  & 000000000100 & 1 & 0.10 & 100010001000 & 1 & 0.10 \\
100110000101 & 3   & 0.30  & 011001010101 & 1 & 0.10 & 000000011011 & 1 & 0.10 \\
111000000101 & 3   & 0.30  & 011101001010 & 1 & 0.10 & 010110000010 & 1 & 0.10 \\
001000000000 & 3   & 0.30  & 000001000000 & 1 & 0.10 & 101010000101 & 1 & 0.10 \\
111110010101 & 2   & 0.20  & 111011000001 & 1 & 0.10 & 000110001000 & 1 & 0.10 \\
011101001100 & 2   & 0.20  & 000000011010 & 1 & 0.10 & 010000000011 & 1 & 0.10 \\
000100000101 & 2   & 0.20  & 011101010000 & 1 & 0.10 & 100110001011 & 1 & 0.10 \\
011101000110 & 2   & 0.20  & 111001000110 & 1 & 0.10 & 111001001010 & 1 & 0.10 \\
011101000001 & 2   & 0.20  & 111101000000 & 1 & 0.10 & 100000001011 & 1 & 0.10 \\
011011000000 & 2   & 0.20  & 111010110100 & 1 & 0.10 & 111011000000 & 1 & 0.10 \\
000001010101 & 2   & 0.20  & 111101000001 & 1 & 0.10 & 010110110101 & 1 & 0.10 \\
000111000001 & 2   & 0.20  & 100000001101 & 1 & 0.10 & 011011000010 & 1 & 0.10 \\
011111000001 & 2   & 0.20  & 100000111111 & 1 & 0.10 & 000000000001 & 1 & 0.10 \\
011001000001 & 2   & 0.20  & 000100000000 & 1 & 0.10 & 000110001010 & 1 & 0.10 \\
100110000001 & 2   & 0.20  & 100110001101 & 1 & 0.10 & 011011011110 & 1 & 0.10 \\
010000000001 & 2   & 0.20  & 000100000001 & 1 & 0.10 & 010011000011 & 1 & 0.10 \\
011100000000 & 2   & 0.20  & 101001110101 & 1 & 0.10 & 100000010101 & 1 & 0.10 \\
111101010101 & 1   & 0.10  & 111111001000 & 1 & 0.10 & 011011001010 & 1 & 0.10 \\
100010000100 & 1   & 0.10  & 111110000001 & 1 & 0.10 & 100010100000 & 1 & 0.10 \\
000001000001 & 1   & 0.10  & 111011001111 & 1 & 0.10 & 100110001111 & 1 & 0.10 \\
110111011111 & 1   & 0.10  & 101111000011 & 1 & 0.10 & 000000001000 & 1 & 0.10 \\
\bottomrule
\end{tabular}
\caption{\centering Exhaustive measurement statistics for the step operator $O\ket{0000}$ utilizing Iceberg QED codes with 3 rounds of mid-circuit error detection ($10^3$ total shots, out of which $872$ successfully passed the QED validation). The bitstring order is $q_t \, x_1 \, x_0 \, y_1 \, y_0 \, q_b \, s_{X,1} \, s_{Z,1} \, s_{X,2} \, s_{Z,2} \, s_{X,3} \, s_{Z,3}$. Here, $x$ and $y$ designate the initial and target logical registers, $q_t$ and $q_b$ represent the top and bottom parity qubits, and $s_{X,i}$ and $s_{Z,i}$ correspond to the $X$ and $Z$ mid-circuit QED syndrome measurement outcomes for round $i \in \{1, 2, 3\}$.}
\label{tab:srw_order1_qed3}
\end{table*}

\begin{table*}[h!]
\centering
\scriptsize
\setlength{\tabcolsep}{2pt}
\begin{tabular}{lrr @{\hspace{0.2cm}} lrr @{\hspace{0.2cm}} lrr}
\toprule
Outcome & Count & Prob. (\%) & Outcome & Count & Prob. (\%) & Outcome & Count & Prob. (\%) \\
\midrule
0000000000000000 & 204 & 20.40 & 1111110000000000 & 203 & 20.30 & 0111010000000000 & 110 & 11.00 \\
1000100000000000 & 109 & 10.90 & 0001100000000000 & 103 & 10.30 & 1110010000000000 & 98  & 9.80  \\
1111110000101010 & 5   & 0.50  & 0111010000001010 & 4   & 0.40  & 0000000000001000 & 2   & 0.20  \\
0000000010101010 & 2   & 0.20  & 0001100000000010 & 2   & 0.20  & 0001100001101010 & 2   & 0.20  \\
0110110000000110 & 2   & 0.20  & 0111010000000001 & 2   & 0.20  & 0111010000000010 & 2   & 0.20  \\
0111110001010101 & 2   & 0.20  & 1000000000000001 & 2   & 0.20  & 1000110000000000 & 2   & 0.20  \\
1101000001010101 & 2   & 0.20  & 1110010000001010 & 2   & 0.20  & 1110010000101010 & 2   & 0.20  \\
1111010000000001 & 2   & 0.20  & 1111010000111111 & 2   & 0.20  & 1111111010101010 & 2   & 0.20  \\
0000000000000001 & 1   & 0.10  & 0000000000001010 & 1   & 0.10  & 0000000000100000 & 1   & 0.10  \\
0000000110101010 & 1   & 0.10  & 0000010000001101 & 1   & 0.10  & 0000100000000000 & 1   & 0.10  \\
0000100000000001 & 1   & 0.10  & 0000100000000101 & 1   & 0.10  & 0000100000001111 & 1   & 0.10  \\
0000100000111111 & 1   & 0.10  & 0000100001010101 & 1   & 0.10  & 0000100011000101 & 1   & 0.10  \\
0000100011010101 & 1   & 0.10  & 0001000000000000 & 1   & 0.10  & 0001000000011101 & 1   & 0.10  \\
0001000000101110 & 1   & 0.10  & 0001010000000000 & 1   & 0.10  & 0001100000001000 & 1   & 0.10  \\
0001100000001011 & 1   & 0.10  & 0001100000101010 & 1   & 0.10  & 0001100010101010 & 1   & 0.10  \\
0001100100000000 & 1   & 0.10  & 0001101010101010 & 1   & 0.10  & 0001110000000000 & 1   & 0.10  \\
0001110000000001 & 1   & 0.10  & 0001110000001101 & 1   & 0.10  & 0001110000111111 & 1   & 0.10  \\
0010000101010101 & 1   & 0.10  & 0010100001010100 & 1   & 0.10  & 0010110001010101 & 1   & 0.10  \\
0011010000000011 & 1   & 0.10  & 0011010000001001 & 1   & 0.10  & 0011100000000010 & 1   & 0.10  \\
0100000000000101 & 1   & 0.10  & 0101000010101010 & 1   & 0.10  & 0101010000010101 & 1   & 0.10  \\
0101010000110101 & 1   & 0.10  & 0101100000000101 & 1   & 0.10  & 0110010000010101 & 1   & 0.10  \\
0110010101010101 & 1   & 0.10  & 0110011101010101 & 1   & 0.10  & 0110100000000011 & 1   & 0.10  \\
0110110000000001 & 1   & 0.10  & 0110110000000010 & 1   & 0.10  & 0110110000111111 & 1   & 0.10  \\
0111000000000001 & 1   & 0.10  & 0111000000000011 & 1   & 0.10  & 0111000000001101 & 1   & 0.10  \\
0111010000000011 & 1   & 0.10  & 0111010000001101 & 1   & 0.10  & 0111010000100000 & 1   & 0.10  \\
0111010010000000 & 1   & 0.10  & 0111010011000001 & 1   & 0.10  & 0111110000000000 & 1   & 0.10  \\
0111110010001010 & 1   & 0.10  & 0111110101010101 & 1   & 0.10  & 1000000000000000 & 1   & 0.10  \\
1000000000010101 & 1   & 0.10  & 1000000001010101 & 1   & 0.10  & 1000000011111111 & 1   & 0.10  \\
1000011000000000 & 1   & 0.10  & 1000100000000010 & 1   & 0.10  & 1000100000001001 & 1   & 0.10  \\
1000100000001010 & 1   & 0.10  & 1000100000100000 & 1   & 0.10  & 1000100000101010 & 1   & 0.10  \\
1000100100000000 & 1   & 0.10  & 1000100100000100 & 1   & 0.10  & 1000100110101010 & 1   & 0.10  \\
1000110000000001 & 1   & 0.10  & 1000110000000111 & 1   & 0.10  & 1000110000111111 & 1   & 0.10  \\
1000110011111111 & 1   & 0.10  & 1001000000000001 & 1   & 0.10  & 1001000000000100 & 1   & 0.10  \\
1001000000001010 & 1   & 0.10  & 1001000001101010 & 1   & 0.10  & 1001000011100101 & 1   & 0.10  \\
1001000110101001 & 1   & 0.10  & 1001010000000011 & 1   & 0.10  & 1001010000000101 & 1   & 0.10  \\
1001010101010101 & 1   & 0.10  & 1001100000000000 & 1   & 0.10  & 1001100001010101 & 1   & 0.10  \\
1001101001010101 & 1   & 0.10  & 1010100000000100 & 1   & 0.10  & 1010100000001111 & 1   & 0.10  \\
1010100000010110 & 1   & 0.10  & 1010100000111111 & 1   & 0.10  & 1010100001010101 & 1   & 0.10  \\
1010100101010101 & 1   & 0.10  & 1010110000000000 & 1   & 0.10  & 1011000000000111 & 1   & 0.10  \\
1011000000011111 & 1   & 0.10  & 1100100000001010 & 1   & 0.10  & 1101000101010101 & 1   & 0.10  \\
1101001100011001 & 1   & 0.10  & 1101110001110110 & 1   & 0.10  & 1110000000010101 & 1   & 0.10  \\
1110000011010101 & 1   & 0.10  & 1110010000000001 & 1   & 0.10  & 1110010000000010 & 1   & 0.10  \\
1110010000001000 & 1   & 0.10  & 1110010000010000 & 1   & 0.10  & 1110010000100000 & 1   & 0.10  \\
1110010001000000 & 1   & 0.10  & 1110010110101001 & 1   & 0.10  & 1110011010101010 & 1   & 0.10  \\
1110100000000000 & 1   & 0.10  & 1110101010101010 & 1   & 0.10  & 1110110000010101 & 1   & 0.10  \\
1110110000101010 & 1   & 0.10  & 1111010000000101 & 1   & 0.10  & 1111010001110101 & 1   & 0.10  \\
1111010001111111 & 1   & 0.10  & 1111100000000001 & 1   & 0.10  & 1111110000000100 & 1   & 0.10  \\
1111110000001000 & 1   & 0.10  & 1111110000011010 & 1   & 0.10  & 1111110000011110 & 1   & 0.10  \\
1111110000111100 & 1   & 0.10  & 1111110010100000 & 1   & 0.10  & 1111110010101010 & 1   & 0.10  \\
1111110100000000 & 1   & 0.10  & 1111110101000000 & 1   & 0.10  & 1111110110101010 & 1   & 0.10  \\
1111111000000000 & 1   & 0.10  &                  &     &       &                  &     &       \\
\bottomrule
\end{tabular}
\caption{\centering Exhaustive measurement statistics for the step operator $O\ket{0000}$ utilizing Iceberg QED codes with 5 rounds of mid-circuit error detection ($10^3$ total shots, out of which $838$ successfully passed the QED validation). The bitstring order is $q_t \, x_1 \, x_0 \, y_1 \, y_0 \, q_b \, s_{X,1} \, s_{Z,1} \dots s_{X,5} \, s_{Z,5}$. Here, $x$ and $y$ designate the initial and target logical registers, $q_t$ and $q_b$ represent the top and bottom parity qubits, and $s_{X,i}$ and $s_{Z,i}$ correspond to the $X$ and $Z$ mid-circuit QED syndrome measurement outcomes for round $i \in \{1, 2, 3, 4, 5\}$.}
\label{tab:srw_order1_qed5}
\end{table*}

\end{document}